\documentclass[aps,preprint]{revtex4}%
\usepackage{amssymb}
\usepackage{amsfonts}
\usepackage{amsmath}
\usepackage{graphicx}%
\providecommand{\U}[1]{\protect\rule{.1in}{.1in}}

\def\A{\mathcal{A}}
\def\H{\mathcal{H}}
\def\pert{{\rm Pert}}
\def\U{\mathcal{U}}
\newcommand{\ie}{{\it i.e.\/}\ }

\begin{document}
\preprint{ }
\title[Beyond SSM]{Beyond the Spectral Standard Model: Emergence of\\Pati-Salam Unification}
\author{Ali H. Chamseddine$^{1,3}$, Alain Connes$^{2,3,4}$ and Walter D. van
Suijlekom$^{5}$}
\email{chams@aub.edu.lb, alain@connes.org, waltervs@math.ru.nl}
\affiliation{$^{1}$Physics Department, American University of Beirut, Lebanon}
\affiliation{$^{2}$College de France, 3 rue Ulm, F75005, Paris, France}
\affiliation{$^{3}$I.H.E.S. F-91440 Bures-sur-Yvette, France}
\affiliation{$^{4}$Department of Mathematics, The Ohio State University, Columbus OH 43210 USA}
\affiliation{$^{5}$Institute for Mathematics, Astrophysics and Particle Physics, Radboud
University Nijmegen, Heyendaalseweg 135, 6525 AJ Nijmegen, The Netherlands}
\keywords{Noncommutative Geometry, Spectral Action, Standard Model}
\pacs{PACS numbers: 04.62.+v. 02.40.-k, 11.15.-q, 11.30.Ly}

\begin{abstract}
The assumption that space-time is a noncommutative space formed as a product
of a continuous four dimensional manifold times a finite space predicts,
almost uniquely, the Standard Model with all its fermions, gauge fields, Higgs
field and their representations. A strong restriction on the noncommutative
space results from the first order condition which came from the requirement
that the Dirac operator is a differential operator of order one. Without this
restriction, invariance under inner automorphisms requires the inner
fluctuations of the Dirac operator to contain a quadratic piece expressed in
terms of the linear part. We apply the classification of product
noncommutative spaces without the first order condition and show that this
leads immediately to a Pati-Salam $SU(2)_{R}\times SU(2)_{L}\times SU(4)$ type
model which unifies leptons and quarks in four colors. Besides the gauge
fields, there are $16$ fermions in the $(2,1,4)+(1,2,4)$ representation, fundamental
Higgs fields in the $(2,2,1)$, $(2,1,4)$ and $(1,1,1+15)$ representations.
Depending on the precise form of 
the initial Dirac operator 
there are additional Higgs fields which are either composite
depending on the fundamental Higgs fields listed above, or are fundamental
themselves. These additional Higgs fields break spontaneously the Pati-Salam
symmetries at high energies to those of the Standard Model.

\end{abstract}
\maketitle

\section{\bigskip Introduction}

Noncommutative geometry was shown to provide a promising framework for
unification of all fundamental interactions including gravity \cite{C96},
\cite{cc1}, \cite{cc2}, \cite{mc2}, \cite{framework}. Historically, the search
to identify the structure of the noncommutative space followed the bottom-up
approach where the known spectrum of the fermionic particles was used to
determine the geometric data that defines the space. This bottom-up approach
involved an interesting interplay with experiments. While at first the
experimental evidence of neutrino oscillations contradicted the first attempt
\cite{cc2}, it was realized several years later in 2006 (\cite{mc2}) that the
obstruction to get neutrino oscillations was naturally eliminated by dropping
the equality between the metric dimension of space-time (which is equal to $4$
as far as we know) and its $KO$-dimension which is only defined modulo $8$.
When the latter is set equal to $2$ modulo $8$ \ \cite{Barrett},
\cite{CoSMneu} (using the freedom to adjust the geometry of the finite space
encoding the fine structure of space-time) everything works fine, the neutrino
oscillations are there as well as the see-saw mechanism which appears for free
as an unexpected bonus. Incidentally, this also solved the fermionic doubling
problem by allowing a simultaneous Weyl-Majorana condition on the fermions to
halve the degrees of freedom.

The second interplay with experiments occurred a bit later when it became
clear that the mass of the Brout-Englert-Higgs boson would not comply with the
restriction (that $m_{H}\succeq170$ Gev) imposed by the validity of the
Standard Model up to the unification scale. This obstruction to lower $m_{H}$
was overcome in \cite{ccres} simply by taking into account a scalar field
which was already present in the full model which we had computed previously
in \cite{framework}. One lesson which we learned on that occasion is that we
have to take all the fields of the noncommutative spectral model seriously,
without making assumptions not backed up by valid analysis, especially because
of the almost uniqueness of the Standard Model (SM) in the noncommutative setting.

The SM continues to conform to all experimental data. The question remains
whether this model will continue to hold at much higher energies, or whether
there is a unified theory whose low-energy limit is the SM. One indication
that there must be a new higher scale that effects the low energy sector is
the small mass of the neutrinos which is explained through the see-saw
mechanism with a Majorana mass of at least of the order of $10^{11}$Gev. In
addition and as noted above, a scalar field which acquires a vev generating
that mass scale can stabilize the Higgs coupling and prevent it from becoming
negative at higher energies and thus make it consistent with the low Higgs
mass of $126$ Gev \cite{ccres}. Another indication of the need to modify the
SM at high energies is the failure (by few percent) of the three gauge
couplings to be unified at some high scale which indicates that it may be
necessary to add other matter couplings to change the slopes of the running of
the RG\ equations.

This leads us to address the issue of the breaking from the natural algebra
$\mathcal{A}$ which results from the classification of irreducible finite
geometries of $KO$-dimension $6$ (modulo $8$) performed in \cite{CC07b}, to
the algebra corresponding to the SM. This breaking was effected in
\cite{CC07b}, \cite{Conceptual} using the requirement of the first order
condition on the Dirac operator. The first order condition is the requirement
that the Dirac operator is a derivation of the algebra $\mathcal{A}$ into the
commutant of $\hat{\mathcal{A}}=J\mathcal{A}J^{-1}$ where $J$ is the charge
conjugation operator. This in turn guarantees the gauge invariance and
linearity of the inner fluctuations \cite{inner} under the action of the gauge
group given by the unitaries $U=uJuJ^{-1}$ for any unitary $u\in\mathcal{A}$.
This condition was used as a mathematical requirement to select the maximal
subalgebra
\[
\mathbb{C}\oplus\mathbb{H}\oplus M_{3}(\mathbb{C})\subset\mathbb{H}_{R}%
\oplus\mathbb{H}_{L}\oplus M_{4}(\mathbb{C})
\]
which is compatible with the first order condition and is the main reason
behind the unique selection of the SM.

The existence of examples of noncommutative spaces where the first order
condition is not satisfied such as quantum groups and quantum spheres provides
a motive to remove this condition from the classification of noncommutative
spaces compatible with unification \cite{diracalls2q}, \cite{diracs2q},
\cite{diracsu2}, \cite{indexsu2}. This study was undertaken in a companion
paper \cite{CCS} where it was shown that in the general case the inner
fluctuations of $D$ form a semigroup in the product algebra
$\mathcal{A}\otimes\mathcal{A}^{\mathrm{op}}$, and acquire a
quadratic part in addition to the linear part. Physically, this new phenomena
will have an impact on the structure of the Higgs fields which are the
components of the connection along discrete directions. This paper is devoted
to the construction of the physical model that describes the physics beyond
the Standard Model. The methods used build on previous results and derivations
developed over the years. To make this work more accessible we shall attempt
to make the paper self-contained by including the parts needed from previous
works in a brief form.

The plan of this paper is as follows. In section \ref{sect:order1} we review the effect of
removing the first order condition on the form of the inner fluctuations,
emphasizing the semigroup structure. In section \ref{sect:classif} we modify the
classification of irreducible finite geometries in the absence of the first
order condition and show that the resultant algebra is, almost uniquely, given
by $\mathbb{H}_{R}\oplus\mathbb{H}_{L}\oplus M_{4}(\mathbb{C}).$ The model is
then based on a noncommutative geometric space formed as a product of a
continuous four dimensional space times the above discrete space. The
associated connection can be viewed either as a $384\times384$ matrices, or in
more manageable form as the tensor product of matrices. To present the
computations in a comprehensible form that could be checked by others, we give
in section \ref{sect:tensors} a brief review of the tenorial notation we developped before.
We stress that all calculations performed in this article using the tensorial
method are done by hand, but have the advantage that they could also be
checked using algebraic manipulation programs such as Mathematica or Maple. In
section \ref{sect:dirac} we compute the inner fluctuations of the Dirac operator on the
above algebra and determine the field content. In section \ref{sect:sa} we evaluate the
spectral action using a cutoff function and the heat kernel expansion method,
where we \ show that the resultant model is the Pati-Salam \cite{PS}
$SU\left(  2\right)  _{R}\times SU\left(  2\right)  _{L}\times SU\left(
4\right)  $ type model with all the appropriate Higgs fields necessary to
break the symmetry to $U\left(  1\right)  _{\mathrm{em}}\times SU\left(
3\right)  _{\mathrm{c}}.$ In section \ref{sect:sm} we show that this model truncates
correctly to the SM. In section \ref{sect:pot} we analyze the potential and possible
symmetry breaking, noting in particular the novel feature that for certain
initial configurations of the Dirac operator some of the inner fluctuations
represented as Higgs fields are fundamental while others are made of quadratic
products of the fundamental ones. For generic initial Dirac operators all
Higgs fields are fundamental. Section \ref{sect:app} is the appendix where all details of
the calculation are given and where we illustrate the power and precision of
noncommutative geometric methods by showing how all the physical fields arise.
This is done to the benefit of researchers interested in becoming
practitioners in the field.

\section{First-order condition and inner fluctuations}
\label{sect:order1}
We briefly summarize the generalization of inner fluctuations to real spectral triples that fail on the first-order condition, as presented in \cite{CCS}. 
In this case, the usual prescription \cite{C96} does not apply, since the operator $D+A\pm JAJ^{-1}$ with gauge potential $A = \sum_j a_j [D,b_j]$ $(a_j,b_j \in \A)$ does not behave well with respect to the action of the gauge group $\U(\A)$. In fact, one would require that conjugation of the fluctuated Dirac operator by the unitary operator $U:=u Ju J^{-1}$ for $u \in \U(\A)$ can be implemented by a usual type of gauge transformation $A \mapsto A^u = u[D,u^*]+ uAu^*$ so that
$$
D+A\pm JAJ^{-1} \mapsto
U (D+A\pm JAJ^{-1}) U^* \equiv D+A^u \pm JA^uJ^{-1}
$$
However, the simple argument only works if $[JuJ^{-1},A] = 0$ for gauge potentials $A$ of the above form and $u \in \U(\A)$, that is, if the first-order condition is satisfied.

For real spectral triples that possibly fail on the first-order condition one starts with a self-adjoint, {\em universal} one-form
\begin{equation}\label{form}
A = \sum_j a_j \delta(b_j); \qquad (a_j,b_j \in \A).
\end{equation}
The inner fluctuations of a real spectral triple $(\A,\H,D; J)$ are then given by
\begin{align}\label{opDprime}
D' = D+A_{(1)}+\tilde A_{(1)} + A_{(2)}
\end{align}
where
\begin{align*}
A_{(1)} &:= \sum_j a_j [D,b_j],\\
\tilde A_{(1)} &:= \sum_j \hat a_j [D,\hat b_j]; \qquad \hat{a}_i = J a_i J^{-1} , \qquad \hat{b}_i = J b_i J^{-1} ,\\
A_{(2)} &:= \sum_j \hat a_j [ A_{(1)}, \hat b_j] = \sum_{j,k} \hat a_j a_k [ [D,b_k], \hat b_j].
\end{align*}
Clearly $A_{(2)}$ which depends quadratically on the fields in $A_{(1)}$  vanishes when the first order condition is satisfied, thus reducing to the usual formulation of inner fluctuations. As such, we will interpret the terms $A_{(2)}$ as non-linear corrections to the {\em first-order}, linear inner fluctuations $A_{(1)}$ of $(\A,\H,D;J)$. 

The need for such quadratic terms can also be seen from the structure of {\em pure gauge} fluctuations $D \mapsto U D U^*$ with $U = u JuJ^{-1}$ and $u \in \U(\A)$. Indeed, in the absence of the first order condition we find that
$$
UDU^* = u[D,u^*] + \hat u [D,\hat u^*] + \hat u[u[D,u^*], \hat u^*].
$$
In the above prescription this corresponds to taking as a universal one-form $A=u\delta(u^*)$. 

On a fluctuated Dirac operator $D'$ such gauge transformation act in a similar way as $D'  \mapsto U D'  U^*$. 
By construction, it is implemented by the gauge transformation
$$
A \mapsto u A u^* + u \delta(u^*)
$$
in the universal differential calculus. In particular, this implies that
$$
A_{(1)} \mapsto u A_{(1)} u^* + u [D,u^*] 
$$
so the first-order inner fluctuations transform as usual. For the term $A_{(2)}$ we compute that a gauge transformation acts as
$$
A_{(2)} \mapsto JuJ^{-1} A_{(2)} J u^* J^{-1} + JuJ^{-1} [u[D,u^*],Ju^* J^{-1}]
$$
where the $A_{(2)}$ on the right-hand-side is expressed using the gauge transformed $A_{(1)}$. This non-linear gauge transformation for $A_{(2)}$ confirms our interpretation of $A_{(2)}$ as the non-linear contribution to the inner fluctuations.

It turns out \cite{CCS} that inner fluctuations come from the action on operators in Hilbert space of a semi-group $\pert(\A)$ of {\em inner perturbations} which only depends on the involutive algebra $\A$ and extends the unitary group of $\A$. More precisely, the semi-group $\pert(\A)$ consists of {\it normalized self-adjoint} elements in $\A \otimes \A^{\mathrm{op}}$:
$$
\pert(\A) := \left\{ \sum_j a_j \otimes b_j^{\mathrm{op}} \in \A \otimes \A^{\mathrm{op}} : \sum_j a_j b_j = 1, \quad \sum_j  a_j \otimes b_j^{\mathrm{op}} = \sum_j  b_j^* \otimes a_j^{*\mathrm{op}} \right\}
$$
with $\A^{\mathrm{op}}$ the involutive algebra $\A$ but with the opposite product $(ab)^{\mathrm{op}} = b^{\mathrm{op}} a^{\mathrm{op}}$. The semi-group product is inherited from the multiplication in the algebra $\A \otimes \A^{\mathrm{op}}$, that is:
$$
\left(\sum_i a_i \otimes b_i^{\mathrm{op}} \right)
\left(\sum_j a_j' \otimes (b_j')^{\mathrm{op}} \right) = \sum_{i,j} a_i a_j' \otimes (b_j' b_i)^{\mathrm{op}},
$$
which indeed respects the above normalization and self-adjointness condition. 
Note that the unitary group of $\A$ is mapped to $\pert(\A)$ by sending a unitary $u$ to $u \otimes u^{*\mathrm{op}}$. 

Given a spectral triple $(\A,\H,D)$ an inner fluctuation of $D$ by an element $\sum_j a_j \otimes b_j^{\mathrm{op}}$ in $\pert(\A)$ is now simply given by
$$
D \mapsto \sum_j a_j D b_j 
. 
$$ 
This covers both cases of ordinary spectral triples and real spectral triples (\ie those which are equipped with the operator $J$). In the latter case one simply uses the natural homomorphism of semi-groups $\mu:\pert(\A)\to \pert(\A\otimes \hat \A)$ given by $\mu(A)=A\otimes \hat A$. Explicitly, this implies for real spectral triples the following transformation rule:
$$
D \mapsto \sum_{i.j} a_i \hat a_j D b_i \hat b_j 
$$
which can indeed be shown \cite[Proposition 5]{CCS} to coincide with the above \eqref{opDprime}. 

The structure of a semi-group implies in particular that inner fluctuations of inner fluctuations are still inner fluctuations ---a fact which is not at all direct when looking at Equation \eqref{opDprime}--- and that the corresponding algebraic rules are unchanged by passing from ordinary spectral triples to real spectral triples.

\section{Classification of finite geometries without first order condition}
\label{sect:classif}
Some time ago the question of classifying finite noncommutative spaces was
carried out in \cite{CC07b}. The main restriction came from requiring that
spinors which belong to the product of the continuous four dimensional space,
times the finite space must be such that the conjugate spinor is not an
independent field, in order to avoid doubling the fermions. This could only be
achieved when the spinors satisfy both the Majorana and Weyl conditions, which
implies that the $KO$-dimension of the finite space be $6$ (mod $8$).
Consistency with the zeroth order condition
\[
\left[  a,b^{\circ}\right]  =0\,,\quad b^{\circ}=Jb^{\ast}J^{-1},~\forall
a,b\in\mathcal{A}%
\]
(since $\mathcal{A}$ is an involutive algebra this condition is the same if
one replaces $b^{\circ}$ by $\hat{b}=JbJ^{-1}$) restricts the center of the
complexified algebra to be $Z\left(  \mathcal{A}_{\mathbb{C}}\right)
=\mathbb{C\oplus C}.$ The dimension of the Hilbert space is then restricted to
be the square of an integer. The algebra is then of the form%
\[
M_{k}\left(  \mathbb{C}\right)  \oplus M_{k}\left(  \mathbb{C}\right)  .
\]
A symplectic symmetry imposed on the first algebra forces $k$ to be even
$k=2a$ and the algebra to be of quaternionic matrices of the form
$M_{a}\left(  \mathbb{H}\right)  .$ {{The existence of the chirality operator
}}breaks $M_{a}\left(  \mathbb{H}\right)  $ and {{further restricts the
integer }}$a$ {{to be even, and thus the number of fundamental fermions must
be of the form }}$4a^{2}$ where $a$ is an even integer. This shows that the
first possible realistic case is the finite space with $k=4$ to be based on
the algebra
\begin{equation}
\mathcal{A}=\mathbb{H}_{R}\oplus\mathbb{H}_{L}\oplus M_{4}\left(
\mathbb{C}\right)  . \label{algebra}%
\end{equation}
A further restriction arises from the first order condition requiring the
commutation of the commutator $\left[  D,a\right]  $ where $D$ is the Dirac
operator and $a\in\mathcal{A}$ with elements $b^{\circ}$, $b\in\mathcal{A}$,
\[
\left[  \left[  D,a\right]  ,b^{\circ}\right]  =0,\qquad a,b\in\mathcal{A}%
,\qquad b^{\circ}=Jb^{\ast}J^{-1}%
\]
(since $\mathcal{A}$ is an involutive algebra this condition is the same if
one replaces $b^{\circ}$ by $\hat{b}=JbJ^{-1}$) This condition, together with
the requirement that the neutrinos must acquire a Majorana mass restricts the
above algebra further to the subalgebra
\begin{equation}
\mathbb{C}\oplus\mathbb{H}\oplus M_{3}\left(  \mathbb{C}\right)  .
\label{subalg}%
\end{equation}
The question is whether the first order condition is an essential requirement
for noncommutative spaces. There are known examples of noncommutative spaces
where the first order condition is not satisfied such as the quantum group
$SU\left(  2\right)  _{q}$ (\cite{diracsu2}, \cite{indexsu2}). As recalled in the previous section, the main
novelty of not imposing the first order condition is that the fluctuations of
the Dirac operator (gauge and Higgs fields) will not be linear anymore and
part of it $A_{\left(  2\right)  }$ will depend quadratically on the fields
appearing in $A_{\left(  1\right)  }.$ In this work we shall study the
resulting noncommutative space without imposing the first order condition on
the Dirac operator. Our starting point, however, will be an initial Dirac
operator (without fluctuations) satisfying the first order condition relative
to the subalgebra \eqref{subalg}, but inner fluctuations would spoil this property.

The noncommutative geometric setting provided answers to some of the basic
questions about the SM, such as the number of fermions in one family, the
nature of the gauge symmetries and their fields, the fermionic
representations, the Higgs fields as gauge fields along discrete directions,
the phenomena of spontaneous symmetry breaking as well many other explanations
\cite{framework}. In other words, noncommutative geometry successfully gave a
geometric setting for the SM. The dynamics of the model was then determined by
the spectral action principle which is based on the idea that all the
geometric invariants of the space can be found in the spectrum of the Dirac
operator of the associated space. Indeed it was shown that the spectral
action, which is a function of the Dirac operator, can be computed and gives
the action of the SM coupled to gravity valid at some high energy scale. When
the couplings appearing in this action are calculated at low energies by
running the RG equations one finds excellent agreement with all known results
to within few percents.

The first order condition is what restricted a more general gauge symmetry
based on the algebra $\mathbb{H}_{R}\oplus\mathbb{H}_{L}\oplus M_{4}\left(
\mathbb{C}\right)  $ to the subalgebra $\mathbb{C}\oplus\mathbb{H}\oplus
M_{3}\left(  \mathbb{C}\right)  .$ It is thus essential to understand the
physical significance of such a requirement. In what follows we shall examine
the more general algebra allowed without the first order condition, and shall
show that the number of fundamental fermions is still dictated to be $16$. We
determine the inner automorphisms of the algebra $\mathcal{A}$ and show that
the resulting gauge symmetry is a Pati-Salam type left-right model
\[
SU\left(  2\right)  _{R}\times SU\left(  2\right)  _{L}\times SU\left(
4\right)
\]
where $SU\left(  4\right)  $ is the color group with the lepton number as the
fourth color. In addition we observe that the Higgs fields appearing in
$A_{\left(  2\right)  }$ are composite and depend quadratically on those
appearing in $A_{\left(  1\right)  }$ provided that the initial Dirac operator
(without fluctuations) satisfies the order one condition relative to the subalgebra \eqref{subalg}. Otherwise, there will
be additional fundamental Higgs fields. In particular, the representations of
the fundamental Higgs fields when the initial Dirac operator satisfies the
order one condition are $\left(  2_{R},2_{L},1\right)  ,$ $\left(  2_{R}%
,1_{L},4\right)  $ and $\left(  1_{R},1_{L},1+15\right)  $ with respect to
$SU\left(  2\right)  _{R}\times SU\left(  2\right)  _{L}\times SU\left(
4\right)  .$ When such an order one condition is not satisfied for the initial
Dirac operator, the representations of the additional Higgs fields are
$\left(  3_{R},1_{L},10\right)  $, $\left(  1_{R},1_{L},6\right)  $ and
$\left(  2_{R},2_{L},1+15\right)  .$ There are simplifications if the Yukawa
coupling of the up quark is equated with that of the neutrino and of the down
quark equated with that of the electron. In addition the $1+15$ of $SU\left(
4\right)  $ decouple if we assume that at unification scale there is exact
$SU\left(  4\right)  $ symmetry between the quarks and leptons. The resulting
model is very similar to the one considered by Marshak and Mohapatra \cite{MM}.

\section{Summary of tensor notation}
\label{sect:tensors}
Although it is possible to use matrix notation to deal with the physical
model, the fact that the matrix representation (which is a product of
matrices) is $384\times384$ dimensional making the task daunting and not very
transparent, although only involving products of matrices. We find it much
more efficient and practical to use a tensorial notation which simplifies
greatly the algebraic operations. This also has the added advantage of
allowing to check all the steps using computer programs with algebraic
manipulations such as Mathematica and Maple.

We will restrict to the case where {{$Z\left(  \mathcal{A}_{\mathbb{C}%
}\right)  =\mathbb{C\oplus C}.$ }}An element of the Hilbert space $\Psi
\in{\mathcal{H}}$ is represented by
\begin{equation}
\Psi_{M}=\left(
\begin{array}
[c]{c}%
\psi_{A}\\
\psi_{A^{^{\prime}}}%
\end{array}
\right)  ,\quad\psi_{A^{\prime}}=\psi_{A}^{c}%
\end{equation}
where $\psi_{A}^{c}$ is the conjugate spinor to $\psi_{A}.$ Thus all primed
indices $A^{\prime}$ correspond to the Hilbert space of conjugage spinors. It
is acted on by both the left algebra $M_{2}\left(  \mathbb{H}\right)  $ and
the right algebra $M_{4}\left(  \mathbb{C}\right)  $. Therefore the index $A$
can take $16$ values and is represented by
\begin{equation}
A=\alpha I
\end{equation}
where the index $\alpha$ is acted on by quaternionic matrices and the index
$I$ \ by $M_{4}\left(  \mathbb{C}\right)  $ matrices. Moreover, when grading
breaks $M_{2}\left(  \mathbb{H}\right)  $ into $\mathbb{H}_{R}\oplus
\mathbb{H}_{L}$ the index $\alpha$ is decomposed to $\alpha=\overset{.}{a},a$
where $\overset{.}{a}=\overset{.}{1},\overset{.}{2}$ (dotted index) is acted
on by the first quaternionic algebra $\ \mathbb{H}_{R}$ and $a=1,2$ is acted
on by the second quaternionic algebra $\ \mathbb{H}_{L}$ . When $M_{4}\left(
\mathbb{C}\right)  $ breaks into $\mathbb{C}\oplus M_{3}\left(  \mathbb{C}%
\right)  $ (due to symmetry breaking or through the use of the order one
condition) the index $I$ is decomposed into $I=1,i$ where the $1$ is acted on
by the $\mathbb{C}$ and the $i$ by $M_{3}\left(  \mathbb{C}\right)  .$
Therefore the various components of the spinor $\psi_{A}$ are
\begin{align*}
\psi_{\alpha I} &  =\left(
\begin{array}
[c]{cccc}%
\nu_{R} & u_{iR} & \nu_{L} & u_{iL}\\
e_{R} & d_{iR} & e_{L} & d_{iL}%
\end{array}
\right)  \\
&  =\left(  \psi_{\overset{.}{a}1},\psi_{\overset{.}{a}i},\psi_{a1},\psi
_{ai}\right)  ,\qquad a=1,2,\qquad a=\overset{.}{1},\overset{.}{2},\qquad
i=1,2,3.
\end{align*}
The power of the abstract notation can be seen by noting that the Dirac action
takes the very simple form
\begin{equation}
\Psi_{M}^{\ast}D_{M}^{N}\Psi_{N}%
\end{equation}
which could be expanded to give
\begin{equation}
\psi_{A}^{\ast}D_{A}^{B}\psi_{B}+\psi_{A^{\prime}}^{\ast}D_{A^{\prime}}%
^{B}\psi_{B}+\psi_{A}^{\ast}D_{A}^{B^{^{\prime}}}\psi_{B^{^{\prime}\prime}%
}+\psi_{A^{\prime}}^{\ast}D_{A^{\prime}}^{B^{\prime}}\psi_{B^{\prime}}%
\end{equation}
The Dirac operator can be written in matrix form%
\begin{equation}
D=\left(
\begin{array}
[c]{cc}%
D_{A}^{B} & D_{A}^{B^{^{\prime}}}\\
D_{A^{^{\prime}}}^{B} & D_{A^{^{\prime}}}^{B^{^{\prime}}}%
\end{array}
\right)  ,
\end{equation}
where $\quad$%
\begin{align}
A &  =\alpha I,\quad\alpha=1,\cdots,4,\quad I=1,\cdots,4\\
\quad A^{\prime} &  =\alpha^{\prime}I^{\prime},\quad\alpha^{\prime}=1^{\prime
},\cdots,4^{\prime},\quad I=1^{\prime},\cdots,4^{\prime}%
\end{align}
Thus $D_{A}^{B}=D_{\alpha I}^{\beta J}$ . Elements of the algebra
\begin{equation}
\mathcal{A}=M_{4}\left(  \mathbb{C}\right)  \oplus M_{4}\left(  \mathbb{C}%
\right)
\end{equation}
are represented by
\begin{equation}
a=\left(
\begin{array}
[c]{cc}%
X_{\alpha}^{\beta}\delta_{I}^{J} & 0\\
0 & \delta_{\alpha^{\prime}}^{\beta^{\prime}}Y_{I^{\prime}}^{J^{\prime}}%
\end{array}
\right)
\end{equation}
where the first block is the tensor product of elements of $M_{4}\left(
\mathbb{C}\right)  \otimes1_{4}$ and the second blcok is the tensor product of
elements of $1_{4}\otimes M_{4}\left(  \mathbb{C}\right)  .$ The reality
operator $J$ is anti-linear and interchange the first and second blocks and
satsify $J^{2}=1$. It is represented by
\begin{equation}
J=\left(
\begin{array}
[c]{cc}%
0 & \delta_{\alpha}^{\beta^{\prime}}\delta_{I}^{J^{\prime}}\\
\delta_{\alpha^{\prime}}^{\beta}\delta_{I^{\prime}}^{J} & 0
\end{array}
\right)  \times\text{\textrm{complex conjugation}}%
\end{equation}
In this form
\begin{equation}
a^{o}=Ja^{\ast}J^{-1}=\left(
\begin{array}
[c]{cc}%
\delta_{\alpha}^{\beta}Y_{I}^{tJ} & 0\\
0 & X_{\alpha^{\prime}}^{t\beta^{\prime}}\delta_{^{I^{\prime}\prime}%
}^{J^{\prime}}%
\end{array}
\right)
\end{equation}
where the superscript $t$ denotes the transpose matrix. This clearly satisfies
the commutation relation
\begin{equation}
\left[  a,b^{o}\right]  =0.
\end{equation}
Writing
\begin{equation}
b=\left(
\begin{array}
[c]{cc}%
Z_{\alpha}^{\beta}\delta_{I}^{J} & 0\\
0 & \delta_{\alpha^{\prime}}^{\beta^{\prime}}W_{I^{\prime}}^{J^{\prime}}%
\end{array}
\right)
\end{equation}
then
\begin{equation}
b^{o}=\left(
\begin{array}
[c]{cc}%
\delta_{\alpha}^{\beta}W_{I}^{tJ} & 0\\
0 & Z_{\alpha^{\prime}}^{t\beta^{\prime}}\delta_{^{I^{\prime}}}^{J^{\prime}}%
\end{array}
\right)
\end{equation}
and so $\left[  \left[  D,a\right]  ,b^{o}\right]  $ is equal to
\begin{equation}
\left(
\begin{array}
[c]{cc}%
\left[  \left[  D,X\right]  ,W^{t}\right]  _{A}^{B} & \left(  \left(
DY-XD\right)  Z^{t}-W^{t}\left(  DY-XD\right)  \right)  _{A}^{B^{\prime}}\\
\left(  \left(  DX-YD\right)  W^{t}-Z^{t}\left(  DX-YD\right)  \right)
_{A^{\prime}}^{B} & \left[  \left[  D,Y\right]  ,Z^{t}\right]  _{A^{\prime}%
}^{B^{\prime}}%
\end{array}
\right)
\end{equation}
The order one condition is
\begin{equation}
\left[  \left[  D,a\right]  ,b^{o}\right]  =0
\end{equation}
which admits a solution with non-zero mixing between primed and unprimed
indices such as
\begin{equation}
D_{\alpha I}^{\beta^{\prime}K^{\prime}}=\delta_{\alpha}^{\overset{.}{1}}%
\delta_{\overset{.}{1^{\prime}}}^{\beta^{\prime}}\delta_{I}^{1}\delta
_{1^{\prime}}^{K^{\prime}}k^{\ast\nu_{R}}
\end{equation}
only when $a,b$ are restricted to the subalgebra $\mathbb{C} \oplus \mathbb{H} \oplus M_3(\mathbb{C}) \subset \A$. Here the $k^{\ast\nu_{R}}$ are matrices in generation space which will be
assumed to be $3\times3.$ We also note that the property that $DJ=JD$ implies
that
\[
D_{A^{\prime}}^{\quad B^{\prime}}=\overline{D}_{A}^{B}.
\]
We further impose the condition of symplectic isometry on the first
$M_{4}\left(  \mathbb{C}\right)  $
\[
\left(  \sigma_{2}\otimes1\right)  \ \left(  \overline{a}\right)  \left(
\sigma_{2}\otimes1\right)  \ =a,\quad a\in M_{4}\left(  \mathbb{C}\right)
\]
which reduces $M_{4}\left(  \mathbb{C}\right)  $ to $M_{2}\left(
\mathbb{H}\right)  $. From the property of commutation of the grading operator
$G_{\alpha}^{\beta}$ with $M_{2}\left(  \mathbb{H}\right)  $
\[
\left[  G,X\right]  =0
\]
where $G_{\alpha}^{\beta}=\left(
\begin{array}
[c]{cc}%
1_{2} & 0\\
0 & -1_{2}%
\end{array}
\right)  ,$ reduces the algebra $M_{2}\left(  \mathbb{H}\right)  $ to
$\mathbb{H}_{R}\mathbb{\oplus H}_{L}.$ Thus we now have
\[
X_{\alpha}^{\beta}=\left(
\begin{array}
[c]{cc}%
X_{\overset{.}{a}}^{\overset{.}{b}} & 0\\
0 & X_{a}^{b}%
\end{array}
\right)  ,\qquad X_{a}^{b}=\left(
\begin{array}
[c]{cc}%
X_{1}^{1} & X_{1}^{2}\\
-\overline{X}_{1}^{2} & \overline{X}_{1}^{1}%
\end{array}
\right)  \in\mathbb{H}_{L}%
\]
and similarly for $X_{\overset{.}{a}}^{\overset{.}{b}}\in\mathbb{H}_{R}$. In
matrix form the operator $D_{F}$ has the sub-matrices \cite{framework}
\begin{align*}
D_{\alpha1}^{\quad\beta1} &  =\left(
\begin{array}
[c]{cc}%
0 & D_{a1}^{\overset{.}{b}1}\\
D_{\overset{.}{a}1}^{b1} & 0
\end{array}
\right)  ,\qquad D_{a1}^{\overset{.}{b}1}=\left(  D_{\overset{.}{a}1}%
^{b1}\right)  ^{\ast}\equiv D_{a\left(  l\right)  }^{\overset{.}{b}}\\
D_{\alpha i}^{\quad\beta j} &  =\left(
\begin{array}
[c]{cc}%
0 & D_{a\left(  q\right)  }^{\overset{.}{b}}\delta_{i}^{j}\\
D_{\overset{.}{a}\left(  q\right)  }^{b}\delta_{i}^{j} & 0
\end{array}
\right)  ,\qquad D_{\overset{.}{a}\left(  q\right)  }^{b}=\left(  D_{a\left(
q\right)  }^{\overset{.}{b}}\right)  ^{\ast}%
\end{align*}
where
\[
D_{a1}^{\overset{.}{b}1}=D_{a\left(  l\right)  }^{\overset{.}{b}}=\left(
\begin{array}
[c]{cc}%
k^{\ast\nu} & 0\\
0 & k^{\ast e}%
\end{array}
\right)  ,\qquad a=1,2,\quad\overset{.}{b}=\overset{.}{1},\overset{.}{2}%
\]
and%
\[
D_{a\left(  q\right)  }^{\overset{.}{b}}=\left(
\begin{array}
[c]{cc}%
k^{\ast u} & 0\\
0 & k^{\ast d}%
\end{array}
\right)  .
\]
The Yukawa couplings $k^{\nu},$ $k^{e},$ $k^{u},$ $k^{d}$ are $3\times3$
matrices in generation space. Notice that this structure gives Dirac masses to
all the fermions, but Majorana masses only for the right-handed neutrinos.
This was shown in \cite{CC07b} to be the unique possibility consistent with
the first order condition on the subalgebra \eqref{subalg}. We can summarize
all the information about the finite space Dirac operator without
fluctuations, in the tensorial equation
\begin{align}
\label{eq:DFpp}
\left(  D_{F}\right)  _{\alpha I}^{\quad\beta J} &  =\left(  \delta_{\alpha
}^{1}\delta_{\overset{.}{1}}^{\beta}k^{\ast\nu}+\delta_{\alpha}^{\overset
{.}{1}}\delta_{1}^{\beta}k^{\nu}+\delta_{\alpha}^{2}\delta_{\overset{.}{2}%
}^{\beta}k^{\ast e}+\delta_{\alpha}^{\overset{.}{2}}\delta_{2}^{\beta}%
k^{e}\right)  \delta_{I}^{1}\delta_{1}^{J}\\
&  +\left(  \delta_{\alpha}^{1}\delta_{\overset{.}{1}}^{\beta}k^{\ast
u}+\delta_{\alpha}^{\overset{.}{1}}\delta_{1}^{\beta}k^{u}+\delta_{\alpha}%
^{2}\delta_{\overset{.}{2}}^{\beta}k^{\ast d}+\delta_{\alpha}^{\overset{.}{2}%
}\delta_{2}^{\beta}k^{d}\right)  \delta_{I}^{i}\delta_{j}^{J}\delta_{i}^{j} \notag \\
\left(  D_{F}\right)  _{\alpha I}^{\quad\beta^{\prime}K^{\prime}} &
=\delta_{\alpha}^{\overset{.}{1}}\delta_{\overset{.}{1}^{\prime}}%
^{\beta^{\prime}}\delta_{I}^{1}\delta_{1^{\prime}}^{K^{\prime}}k^{\ast\nu_{R}}%
\label{eq:DFppbar}
\end{align}
where $k^{\nu_{R}}$ are Yukawa couplings for the right-handed neutrinos. One
can also consider the special case of lepton and quark unification by
equating
\[
k^{\nu}=k^{u},\qquad k^{e}=k^{d}%
\]
where we expect some simplifications. 

\section{Dirac operator and Inner fluctuations on $\mathbb{H}_{R}%
\oplus\mathbb{H}_{L}\oplus M_{4}\left(  \mathbb{C}\right)  $}
\label{sect:dirac}
Recall that if one considers inner fluctuations of the Dirac operator one finds that the
gauge transformation takes the form
\[
D_{A}\rightarrow UD_{A}U^{\ast},\qquad U=u\,Ju\,J^{-1},\qquad u\in
\mathcal{U}\left(  \mathcal{A}\right)
\]
which implies that
\[
A\rightarrow u\,Au^{\ast}+u\delta\left(  u^{\ast}\right)  .
\]
This in turn gives
\begin{align*}
A_{\left(  1\right)  }  &  \rightarrow uA_{\left(  1\right)  }u^{\ast
}+u\left[  D,u^{\ast}\right]  
\\
A_{\left(  2\right)  }  &  \rightarrow Ju\,J^{-1}A_{\left(  2\right)
}Ju^{\ast}\,J^{-1}+Ju\,J^{-1}\left[  u\left[  D,u^{\ast}\right]  ,Ju\,^{\ast
}J^{-1}\right]
\end{align*}
where the $A_{\left(  2\right)  }$ in the right hand side is computed using
the gauge transformed $A_{\left(  1\right)  }$. Thus $A_{\left(  1\right)  }$
is a one-form and behaves like the usual gauge transformations. On the other
hand $A_{\left(  2\right)  }$ transforms non-linearly and includes terms with
quadratic dependence on the gauge transformations.

We now proceed to compute the Dirac operator on the product space $M\times F$
. The initial operator is given by
\[
D=\gamma^{\mu}D_{\mu}\otimes1+\gamma_{5}D_{F}%
\]
where $\gamma^{\mu}D_{\mu}=\gamma^{\mu}\left(  \partial_{\mu}+\frac{1}%
{4}\omega_{\mu}^{\quad ab}\gamma_{ab}\right)  $ is the Dirac operator on the
four dimensional spin manifold. Then the Dirac operator including inner
fluctuations is given by
\[
D_{A}=D+A_{\left(  1\right)  }+JA_{\left(  1\right)  }J^{-1}+A_{\left(
2\right)  }%
\]%
\begin{align*}
A_{\left(  1\right)  } &  =%
{\displaystyle\sum}
a\left[  D,b\right]  \\
A_{\left(  2\right)  } &  =%
{\displaystyle\sum}
a\left[  JA_{\left(  1\right)  }J^{-1},b\right]  .
\end{align*}
The computation is very involved thus for clarity we shall collect all the
details in the appendix and only quote the results in what follows. The
different components of the operator $D_{A}$ are then given by
\begin{align*}
\left(  D_{A}\right)  _{\overset{.}{a}I}^{\overset{.}{b}J} &  =\gamma^{\mu
}\left(  D_{\mu}\delta_{\overset{.}{a}}^{\overset{.}{b}}\delta_{I}^{J}%
-\frac{i}{2}g_{R}W_{\mu R}^{\alpha}\left(  \sigma^{\alpha}\right)
_{\overset{.}{a}}^{\overset{.}{b}}\delta_{I}^{J}-\delta_{\overset{.}{a}%
}^{\overset{.}{b}}\left(  \frac{i}{2}gV_{\mu}^{m}\left(  \lambda^{m}\right)
_{I}^{^{J}}+\frac{i}{2}gV_{\mu}\delta_{I}^{J}\right)  \right)  \\
\left(  D_{A}\right)  _{aI}^{bJ} &  =\gamma^{\mu}\left(  D_{\mu}\delta_{a}%
^{b}\delta_{I}^{J}-\frac{i}{2}g_{L}W_{\mu L}^{\alpha}\left(  \sigma^{\alpha
}\right)  _{a}^{b}\delta_{I}^{J}-\delta_{a}^{b}\left(  \frac{i}{2}gV_{\mu}%
^{m}\left(  \lambda^{m}\right)  _{I}^{^{J}}+\frac{i}{2}gV_{\mu}\delta_{I}%
^{J}\right)  \right)
\end{align*}
where the fifteen $4\times4$ matrices $\left(  \lambda^{m}\right)  _{I}^{^{J}%
}$ are traceless and generate the group $SU\left(  4\right)  $ and $W_{\mu
R}^{\alpha},$ $W_{\mu L}^{\alpha},$ $V_{\mu}^{m}$ are the gauge fields of
$SU\left(  2\right)  _{R}$, $SU\left(  2\right)  _{L}$, and $SU\left(
4\right)  .$ The requirement that $A$ is unimodular implies that
\[
\mathrm{Tr}\left(  A\right)  =0
\]
which gives the condition
\[
V_{\mu}=0.
\]
In addition we have
\begin{align}
\left(  D_{A}\right)  _{\overset{.}{a}I}^{bJ} &  =\gamma_{5}\left(  \left(
k^{\nu}\phi_{\overset{.}{a}}^{b}+k^{e}\widetilde{\phi}_{\overset{.}{a}}%
^{b}\right)  \Sigma_{I}^{J}+\left(  k^{u}\phi_{\overset{.}{a}}^{b}%
+k^{d}\widetilde{\phi}_{\overset{.}{a}}^{b}\right)  \left(  \delta_{I}%
^{J}-\Sigma_{I}^{J}\right)  \right)  \equiv\gamma_{5}\Sigma_{\overset{.}{a}%
I}^{bJ}\label{dependent higgs}\\
\left(  D_{A}\right)  _{\overset{.}{a}I}^{\overset{.}{b}^{\prime}J^{\prime}}
&  =\gamma_{5}k^{\ast\nu_{R}}\Delta_{\overset{.}{a}J}\Delta_{\overset{.}{b}%
I}\equiv\gamma_{5}H_{\overset{.}{a}I\overset{.}{b}J}\nonumber
\end{align}
where the Higgs field $\phi_{\overset{.}{a}}^{b}$ is in the $\left(
2_{R},\overline{2}_{L},1\right)  $ of the product gauge group $SU\left(
2\right)  _{R}\times SU\left(  2\right)  _{L}\times SU\left(  4\right)  $, and
$\Delta_{\overset{.}{a}J}$ is in the $\left(  2_{R,},1_{L},4\right)  $
representation while $\Sigma_{I}^{J}$ is in the $\left(  1_{R},1_{L}%
,1+15\right)  $ representation. The field $\widetilde{\phi}_{\overset{.}{a}%
}^{b}$ is not an independent field and is given by%
\[
\widetilde{\phi}_{\overset{.}{a}}^{b}=\sigma_{2}\overline{\phi}_{\overset{.}{a}%
}^{b}\sigma_{2}.
\]
Note that the field $\Sigma_{I}^{J}$ decouples (and set to $\delta_{I}%
^{1}\delta_{1}^{J}$ ) in the special case when there is lepton and quark
unification of the couplings
\[
k^{\nu}=k^{u},\qquad k^{e}=k^{d}.
\]
In case when the initial Dirac operator satisfies the order one condition for the subalgebra \eqref{subalg}, then the $A_{\left(  2\right)  }$ part of the connection becomes a composite Higgs field where the Higgs field $\Sigma_{\overset{.}{a}I}^{bJ}$ is formed
out of the products of the fields $\phi_{\overset{.}{a}}^{b}$ and $\Sigma
_{I}^{J}$ while the Higgs field $H_{\overset{.}{a}I\overset{.}{b}J}$ is made
from the product of $\Delta_{\overset{.}{a}J}\Delta_{\overset{.}{b}I}.$ For
generic initial Dirac operators, the field $\left(  A_{\left(  2\right)
}\right)  _{\overset{.}{a}I}^{bJ}$ becomes independent. The fields
$\Sigma_{\overset{.}{a}I}^{bJ}$ and $H_{\overset{.}{a}I\overset{.}{b}J}$ will
then not be defined through equation \ref{dependent higgs} and will be in the
$\left(  2_{R},2_{L},1+15\right)  $ and $\left(  3_{R},1_{L},10\right)
+\left(  1_{R},1_{L},6\right)  $ representations of $SU\left(  2\right)
_{R}\times SU\left(  2\right)  _{L}\times SU\left(  4\right)  .$ In addition,
for generic Dirac operator one also generates the fundamental field $\left(
1,2_{L},4\right)  .$ The fact that inner automorphisms form a semigroup
implies that the cases where the Higgs fields contained in the connections
$A_{\left(  2\right)  }$ are either independent fields or depend quadratically
on the fundamental Higgs fields are disconnected. The interesting question
that needs to be addressed is whether the structure of the connection is
preserved at the quantum level. This investigation must be performed in such a
way as to take into account the noncommutative structure of the space. At any
rate, we have here a clear advantage over grand unified theories which suffers
of having  arbitrary and complicated Higgs representations. In the
noncommutative geometric setting, this problem is now solved by having minimal
representations of the Higgs fields. Remarkably, we note that a very close
model to the one deduced here is the one considered by Marshak and Mohapatra
where the $U\left(  1\right)  $ of the left-right model is identified with the
$B-L$ symmetry. They proposed the same Higgs fields that would result starting
with a generic initial Dirac operator not satisfying the first order
condition. Although the broken generators of the $SU\left(  4\right)  $ gauge
fields can mediate lepto-quark interactions leading to proton decay, it was
shown that in all such types of models with partial unification, the proton is
stable. In addition this type of model arises in the first phase of breaking
of $SO\left(  10\right)  $ to $SU\left(  2\right)  _{R}\times SU\left(
2\right)  _{L}\times SU\left(  4\right)  $ and these have been extensively
studied \cite{BM}. The recent work in \cite{DLM13} considers noncommutative
grand unification based on the $k=8$ algebra $M_{4}\left(  \mathbb{H}\right)
\oplus M_{8}\left(  \mathbb{C}\right)  $ keeping the first order condition.

\section{The Spectral Action for the $SU\left(  2\right)  _{R}\times SU\left(
2\right)  _{L}\times SU\left(  4\right)  $ model}
\label{sect:sa}

Having determined the Dirac operator acting on the Hilbert space of spinors in
terms of the gauge fields of $SU\left(  2\right)  _{R}\times SU\left(
2\right)  _{L}\times SU\left(  4\right)  $ and Higgs fields, some of which are
fundamental while others are composite, the next step is to study the dynamics
of these fields as governed by the spectral action principle. The geometric
invariants of the noncommutative space are encoded in the spectrum of the
Dirac operator $D_{A}$. The bosonic action is given by%
\[
\mathrm{Trace\,}\left(  f\left(  D_{A}/\Lambda\right)  \right)
\]
where $\Lambda$ is some cutoff scale and the function $f$ \ is restricted to
be even and positive. Using heat kernel methods the trace can be expressed in
terms of Seeley-de Witt coefficients $a_{n}:$%
\[
\mathrm{Trace}\,f\left(  D_{A}/\Lambda\right)  =%
{\displaystyle\sum\limits_{n=0}^{\infty}}
F_{4-n}\Lambda^{4-n}a_{n}%
\]
where the function $F$ is defined by $F(u)=f\,(v)$ where $u=v^{2},$ thus
$F(D^{2})=f\,(D)$. We define
\[
f_{k}=%
{\displaystyle\int\limits_{0}^{\infty}}
f\left(  v\right)  v^{k-1}dv,\qquad k>0
\]
then
\begin{align*}
F_{4} &  =%
{\displaystyle\int\limits_{0}^{\infty}}
F(u)udu=2%
{\displaystyle\int\limits_{0}^{\infty}}
f(v)v^{3}dv=2f_{4}\\
F_{2} &  =%
{\displaystyle\int\limits_{0}^{\infty}}
F(u)du=2%
{\displaystyle\int\limits_{0}^{\infty}}
f(v)vdv=2f_{2}\\
F_{0} &  =F(0)=f\,(0)=f_{0}\\
F_{-2n} &  =\left(  -1\right)  ^{n}F^{\left(  n\right)  }\left(  0\right)
=\left[  \left(  -1\right)  ^{n}\left(  \frac{1}{2v}\frac{d}{dv}\right)
^{n}f\right]  \left(  0\right)  \qquad n\geq1.
\end{align*}
Using the same notation and formulas as in reference \cite{framework}, the
first Seeley-de Witt coefficient is
\begin{align*}
a_{0} &  =\frac{1}{16\pi^{2}}%
{\displaystyle\int}
d^{4}x\sqrt{g}\text{Tr}\left(  1\right)  \\
&  =\frac{1}{16\pi^{2}}\left(  4\right)  \left(  32\right)  \left(  3\right)
{\displaystyle\int}
d^{4}x\sqrt{g}\\
&  =\frac{24}{\pi^{2}}%
{\displaystyle\int}
d^{4}x\sqrt{g}%
\end{align*}
where the numerical factors come, respectively, from the traces on the
Clifford algebra, the dimensions of the Hilbert space and number of
generations. The second coefficient is
\[
a_{2}=\frac{1}{16\pi^{2}}%
{\displaystyle\int}
d^{4}x\sqrt{g}\text{\textrm{Tr}}\left(  E+\frac{1}{6}R\right)
\]
where $E$ is a $384\times384$ matrix over Hilbert space of three generations
of spinors, whose components are derived and listed in the appendix. Taking
the various traces we get
\begin{align*}
a_{2} &  =\frac{1}{16\pi^{2}}%
{\displaystyle\int}
d^{4}x\sqrt{g}\left(  \left(  R(-96+64\right)  -8\left(  H_{\overset{.}%
{a}I\overset{.}{c}K}H^{\overset{.}{c}K\overset{.}{a}I}+2\Sigma_{\overset{.}%
{a}I}^{cK}\Sigma_{cK}^{\overset{.}{a}I}\right)  \right)  \\
&  =-\frac{2}{\pi^{2}}%
{\displaystyle\int}
d^{4}x\sqrt{g}\left(  R+\frac{1}{4}\left(  H_{\overset{.}{a}I\overset{.}{c}%
K}H^{\overset{.}{c}K\overset{.}{a}I}+2\Sigma_{\overset{.}{a}I}^{cK}\Sigma
_{cK}^{\overset{.}{a}I}\right)  \right)  .
\end{align*}
It should be understood in the above formula and in what follows, that
whenever the matrices $k^{\nu},k^{u},k^{e},k^{d}$ and $k^{\nu_{R}}$ appear in
an action, one must take the trace over generation space. When the initial
Dirac operator without fluctuations is taken to satisfy the order one
condition, the fields $H_{\overset{.}{a}I\overset{.}{c}K}$ and $\Sigma
_{\overset{.}{a}I}^{cK}$ will become dependent on the fundamental Higgs
fields. In this case, the mass terms can be expressed in terms of the
fundamental Higgs field to give
\[
H_{\overset{.}{a}I\overset{.}{c}K}H^{\overset{.}{c}K\overset{.}{a}%
I}=\left\vert k^{\nu_{R}}\right\vert ^{2}\left(  \Delta_{\overset{.}{a}%
K}\overline{\Delta}^{\overset{.}{a}K}\right)  ^{2}%
\]
and
\begin{align*}
2\Sigma_{\overset{.}{a}I}^{cK}\Sigma_{cK}^{\overset{.}{a}I} &  =2\left(
\left(  \left(  k^{\nu}-k^{u}\right)  \phi_{\overset{.}{a}}^{c}+\left(
k^{e}-k^{d}\right)  \widetilde{\phi}_{\overset{.}{a}}^{c}\right)  \Sigma
_{I}^{K}+\left(  k^{u}\phi_{\overset{.}{a}}^{c}+k^{d}\widetilde{\phi
}_{\overset{.}{a}}^{c}\right)  \delta_{I}^{K}\right)  \\
&  \left(  \left(  \left(  k^{\ast\nu}-k^{\ast u}\right)  \phi_{c}%
^{\overset{.}{a}}+\left(  k^{\ast e}-k^{\ast d}\right)  \widetilde{\phi}%
_{c}^{\overset{.}{a}}\right)  \Sigma_{K}^{I}+\left(  k^{\ast u}\phi
_{c}^{\overset{.}{a}}+k^{\ast d}\widetilde{\phi}_{c}^{\overset{.}{a}}\right)
\delta_{K}^{I}\right)  .
\end{align*}
The next coefficient is%
\[
a_{4}=\frac{1}{16\pi^{2}}%
{\displaystyle\int}
d^{4}x\sqrt{g}\text{\textrm{Tr}}\left(  \frac{1}{360}\left(  5R^{2}-2R_{\mu
\nu}^{2}+2R_{\mu\nu\rho\sigma}^{2}\right)  1+\frac{1}{2}\left(  E^{2}+\frac
{1}{3}RE+\frac{1}{6}\Omega_{\mu\nu}^{2}\right)  \right)
\]
where $\Omega_{\mu\nu}$ is the $384\times384$ curvature matrix of the
connection $\omega_{\mu}$. Using the expressions for the matrices $E$ and
$\Omega_{\mu\nu}$ derived in the appendix, and taking the traces, we get
\begin{align*}
a_{4} &  =\frac{1}{2\pi^{2}}%
{\displaystyle\int}
d^{4}x\sqrt{g}\left[  -\frac{3}{5}C_{\mu\nu\rho\sigma}^{2}+\frac{11}%
{30}R^{\ast}R^{\ast}+g_{L}^{2}\left(  W_{\mu\nu L}^{\alpha}\right)  ^{2}%
+g_{R}^{2}\left(  W_{\mu\nu R}^{\alpha}\right)  ^{2}+g^{2}\left(  V_{\mu\nu
}^{m}\right)  ^{2}\right.  \\
&  \qquad\qquad+\nabla_{\mu}\Sigma_{aI}^{\overset{.}{c}K}\nabla^{\mu}%
\Sigma_{\overset{.}{c}K}^{aI}+\frac{1}{2}\nabla_{\mu}H_{\overset{.}%
{a}I\overset{.}{b}J}\nabla^{\mu}H^{\overset{.}{a}I\overset{.}{b}J}+\frac
{1}{12}R\left(  H_{\overset{.}{a}I\overset{.}{c}K}H^{\overset{.}{c}%
K\overset{.}{a}I}+2\Sigma_{\overset{.}{a}I}^{cK}\Sigma_{cK}^{\overset{.}{a}%
I}\right)  \\
&  \qquad\qquad\left.  +\frac{1}{2}\left\vert H_{\overset{.}{a}I\overset{.}%
{c}K}H^{\overset{.}{c}K\overset{.}{b}J}\right\vert ^{2}+2H_{\overset{.}%
{a}I\overset{.}{c}K}\Sigma_{bJ}^{\overset{.}{c}K}H^{\overset{.}{a}I\overset
{.}{d}L}\Sigma_{\overset{.}{d}L}^{bJ}+\Sigma_{aI}^{\overset{.}{c}K}%
\Sigma_{\overset{.}{c}K}^{bJ}\Sigma_{bJ}^{\overset{.}{d}L}\Sigma_{\overset
{.}{d}L}^{aI}\right]
\end{align*}
where $C_{\mu\nu\rho\sigma}$ is the Weyl tensor. Thus the bosonic spectral
action to second order is given by%
\[
S=F_{4}\Lambda^{4}a_{0}+F_{2}\Lambda^{2}a_{2}+F_{0}a_{4}+\cdots
\]
which finally gives%
\begin{align*}
S_{\mathrm{b}} &  =\frac{24}{\pi^{2}}F_{4}\Lambda^{4}%
{\displaystyle\int}
d^{4}x\sqrt{g}\\
&  -\frac{2}{\pi^{2}}F_{2}\Lambda^{2}%
{\displaystyle\int}
d^{4}x\sqrt{g}\left(  R+\frac{1}{4}\left(  H_{\overset{.}{a}I\overset{.}{c}%
K}H^{\overset{.}{c}K\overset{.}{a}I}+2\Sigma_{\overset{.}{a}I}^{cK}\Sigma
_{cK}^{\overset{.}{a}I}\right)  \right)  \\
&  +\frac{1}{2\pi^{2}}F_{0}%
{\displaystyle\int}
d^{4}x\sqrt{g}\left[  \frac{1}{30}\left(  -18C_{\mu\nu\rho\sigma}%
^{2}+11R^{\ast}R^{\ast}\right)  +g_{L}^{2}\left(  W_{\mu\nu L}^{\alpha
}\right)  ^{2}+g_{R}^{2}\left(  W_{\mu\nu R}^{\alpha}\right)  ^{2}%
+g^{2}\left(  V_{\mu\nu}^{m}\right)  ^{2}\right.  \\
&  +\ \nabla_{\mu}\Sigma_{aI}^{\overset{.}{c}K}\nabla^{\mu}\Sigma_{\overset
{.}{c}K}^{aI}+\frac{1}{2}\nabla_{\mu}H_{\overset{.}{a}I\overset{.}{b}J}%
\nabla^{\mu}H^{\overset{.}{a}I\overset{.}{b}J}+\frac{1}{12}R\left(
H_{\overset{.}{a}I\overset{.}{c}K}H^{\overset{.}{c}K\overset{.}{a}I}%
+2\Sigma_{\overset{.}{a}I}^{cK}\Sigma_{cK}^{\overset{.}{a}I}\right)  \\
&  \left.  +\frac{1}{2}\left\vert H_{\overset{.}{a}I\overset{.}{c}%
K}H^{\overset{.}{c}K\overset{.}{b}J}\right\vert ^{2}+2H_{\overset{.}%
{a}I\overset{.}{c}K}\Sigma_{bJ}^{\overset{.}{c}K}H^{\overset{.}{a}I\overset
{.}{d}L}\Sigma_{\overset{.}{d}L}^{bJ}+\Sigma_{aI}^{\overset{.}{c}K}%
\Sigma_{\overset{.}{c}K}^{bJ}\Sigma_{bJ}^{\overset{.}{d}L}\Sigma_{\overset
{.}{d}L}^{aI}\right]  .
\end{align*}
The physical content of this action is a cosmological constant term, the
Einstein Hilbert term $R,$ a Weyl tensor square term $C_{\mu\nu\rho\sigma}%
^{2},$ kinetic terms for the $SU\left(  2\right)  _{R}\times SU\left(
2\right)  _{L}\times SU\left(  4\right)  $ gauge fields, kinetic terms for the
composite Higgs fields $H_{\overset{.}{a}I\overset{.}{b}J}$ and $\Sigma
_{bJ}^{\overset{.}{c}K}$ as well as mass terms and quartic terms for the Higgs
fields. This is a grand unified Pati-Salam type model with a completely fixed
Higgs structure which we expect to spontaneously break at very high energies
to the $U\left(  1\right)  \times SU\left(  2\right)  \times SU\left(
3\right)  $ symmetry of the SM. We also notice that this action gives the
gauge coupling unification
\[
g_{R}=g_{L}=g.
\]
A test of this model is to check whether this relation when run using RG
equations would give values consistent with the values of the gauge couplings
for electromagnetic, weak and strong interactions at the scale of the $Z$
-boson mass. Having determined the full Dirac operators, including
fluctuations, we can write all the fermionic interactions including the ones
with the gauge vectors and Higgs scalars. It is given by
\begin{align*}
&
{\displaystyle\int}
d^{4}x\sqrt{g}\left\{  \psi_{\overset{.}{a}I}^{\ast}\gamma^{\mu}\left(
D_{\mu}\delta_{\overset{.}{a}}^{\overset{.}{b}}\delta_{I}^{J}-\frac{i}{2}%
g_{R}W_{\mu R}^{\alpha}\left(  \sigma^{\alpha}\right)  _{\overset{.}{a}%
}^{\overset{.}{b}}\delta_{I}^{J}-\delta_{\overset{.}{a}}^{\overset{.}{b}%
}\left(  \frac{i}{2}gV_{\mu}^{m}\left(  \lambda^{m}\right)  _{I}^{^{J}}%
+\frac{i}{2}gV_{\mu}\delta_{I}^{J}\right)  \right)  \psi_{\overset{.}{b}%
J}\right.  \\
&  \qquad\qquad+\psi_{aI}^{\ast}\gamma^{\mu}\left(  D_{\mu}\delta_{a}%
^{b}\delta_{I}^{J}-\frac{i}{2}g_{L}W_{\mu L}^{\alpha}\left(  \sigma^{\alpha
}\right)  _{a}^{b}\delta_{I}^{J}-\delta_{a}^{b}\left(  \frac{i}{2}gV_{\mu}%
^{m}\left(  \lambda^{m}\right)  _{I}^{^{J}}+\frac{i}{2}gV_{\mu}\delta_{I}%
^{J}\right)  \right)  \psi_{bJ}\\
&  \qquad\qquad\left.  +\psi_{\overset{.}{a}I}^{\ast}\gamma_{5}\Sigma
_{\overset{.}{a}I}^{bJ}\psi_{bJ}+\psi_{aI}^{\ast}\gamma_{5}\Sigma
_{aI}^{\overset{.}{b}J}\psi_{\overset{.}{b}J}+C\psi_{\overset{.}{a}I}%
\gamma_{5}H^{\overset{.}{a}I\overset{.}{b}J}\psi_{\overset{.}{b}%
J}+\mathrm{h.c}\right\}
\end{align*}

\section{Truncation to the Standard Model}
\label{sect:sm}
It is easy to see that this model truncates to the Standard Model. The Higgs
field $\phi_{\overset{.}{a}}^{b}$ $=\left(  2_{R},2_{L},1\right)  $ must be
truncated to the Higgs doublet $H$ by writing
\begin{equation*}
\phi_{\overset{.}{a}}^{b}=\delta_{\overset{.}{a}}^{\overset{.}{1}}%
\epsilon^{bc}H_{c}.%
\end{equation*}
The other Higgs field $\Delta_{\overset{.}{a}I}=\left(  2_{R},1,4\right)  $ is
truncated to a real singlet scalar field
\begin{equation*}
\Delta_{\overset{.}{a}I}=\delta_{\overset{.}{a}}^{\overset{.}{1}}\delta
_{I}^{1}\sqrt{\sigma}.%
\end{equation*}
These then imply the relations%
\begin{align*}
\Sigma_{\overset{.}{a}I}^{bJ}  &  =\left(  \delta_{\overset{.}{a}%
}^{\overset{.}{1}}k^{\nu}\epsilon^{bc}H_{c}+\delta_{\overset{.}{a}%
}^{\overset{.}{2}}\overline{H}^{b}k^{e}\right)  \delta_{I}^{1}\delta_{1}%
^{J}+\left(  \delta_{\overset{.}{a}}^{\overset{.}{1}}k^{u}\epsilon^{bc}%
H_{c}+\delta_{\overset{.}{a}}^{\overset{.}{2}}k^{d}\overline{H}^{b}\right)
\delta_{I}^{i}\delta_{j}^{J}\delta_{i}^{j}\\
H_{\overset{.}{a}I\overset{.}{b}J}  &  =\delta_{\overset{.}{a}}%
^{\overset{.}{1}}\delta_{\overset{.}{b}}^{\overset{.}{1}}k^{\nu_{R}}\delta
_{I}^{1}\delta_{1}^{J}\sigma\\
g_{R}W_{\mu R}^{3}  &  =g_{1}B_{\mu},\qquad W_{\mu R}^{\pm}=0\\
\sqrt{\frac{3}{2}}gV_{\mu}^{15}  &  =-g_{1}B_{\mu}\qquad\left(  V_{\mu
}\right)  _{1}^{i}=0
\end{align*}
where $V_{\mu}^{15}$ is the $SU(4)$ gauge field corresponding to the
generator
\begin{equation*}
\lambda^{15}=\frac{1}{\sqrt{6}}\mathrm{diag}\left(  3,-1,-1,-1\right)
\end{equation*}
which could be identified with the $B-L$ generator. In particular the
components $\left(  D_{A}\right)  _{\overset{.}{1}1}^{\overset{.}{1}1}$ and
$\left(  D_{A}\right)  _{\overset{.}{2}1}^{\overset{.}{2}1}$ of the Dirac
operator simplify to
\begin{align*}
\left(  D_{A}\right)  _{\overset{.}{1}1}^{\overset{.}{1}1}  &  =\gamma^{\mu
}\left(  D_{\mu}-\frac{i}{2}g_{R}W_{\mu R}^{\alpha}\left(  \sigma^{\alpha
}\right)  _{\overset{.}{1}}^{\overset{.}{1}}-\left(  \frac{i}{2}gV_{\mu}%
^{m}\left(  \lambda^{m}\right)  _{1}^{^{1}}\right)  \right) \nonumber\\
&  =\gamma^{\mu}\left(  D_{\mu}-\frac{i}{2}g_{R}W_{\mu R}^{3}-\left(  \frac
{i}{2}gV_{\mu}^{15}\sqrt{\frac{3}{2}}\right)  \right) \\
&  =\gamma^{\mu}D_{\mu}%
\end{align*}%
\begin{align*}
\left(  D_{A}\right)  _{\overset{.}{2}1}^{\overset{.}{2}1}  &  =\gamma^{\mu
}\left(  D_{\mu}-\frac{i}{2}g_{R}W_{\mu R}^{\alpha}\left(  \sigma^{\alpha
}\right)  _{\overset{.}{2}}^{\overset{.}{2}}-\left(  \frac{i}{2}gV_{\mu}%
^{m}\left(  \lambda^{m}\right)  _{1}^{^{1}}\right)  \right) \nonumber\\
&  =\gamma^{\mu}\left(  D_{\mu}+\frac{i}{2}g_{R}W_{\mu R}^{3}-\left(  \frac
{i}{2}gV_{\mu}^{15}\sqrt{\frac{3}{2}}\right)  \right) \\
&  =\gamma^{\mu}\left(  D_{\mu}+ig_{1}B_{\mu}\right)
\end{align*}
which are identified with the Dirac operators acting on the right-handed
neutrino and right-handed electron. Similar substitutions give the action of
the Dirac operators on the remaining fermions and give the expected results.
We now compute the various terms in the spectral action. First for the mass
terms we have
\begin{align*}
\frac{1}{4}H_{\overset{.}{a}I\overset{.}{b}J}H^{\overset{.}{b}J\overset{.}{a}%
I}  &  =\frac{1}{4}\left(  \delta_{\overset{.}{a}}^{1}\delta_{\overset{.}{b}%
}^{1}k^{\nu_{R}}\delta_{I}^{1}\delta_{1}^{J}\sigma\right)  \left(  \delta
_{1}^{\overset{.}{a}}\delta_{1}^{\overset{.}{b}}\delta_{1}^{I}\delta_{1}%
^{J}k^{\ast\nu_{R}}\sigma\right) \nonumber\\
&  =\frac{1}{4}\mathrm{tr}\left\vert k^{\nu_{R}}\right\vert ^{2}\sigma
^{2}=\frac{1}{4}c\sigma^{2}\\
\frac{1}{2}\Sigma_{\overset{.}{a}I}^{cK}\Sigma_{cK}^{\overset{.}{a}I}  &
=\frac{1}{2}\left\vert \left(  \delta_{\overset{.}{a}}^{\overset{.}{1}}k^{\nu
}\epsilon^{bc}H_{c}+\delta_{\overset{.}{a}}^{\overset{.}{2}}\overline{H}%
^{b}k^{e}\right)  \delta_{I}^{1}\delta_{1}^{J}+\left(  \delta_{\overset{.}{a}%
}^{\overset{.}{1}}k^{u}\epsilon^{bc}H_{c}+\delta_{\overset{.}{a}%
}^{\overset{.}{2}}k^{d}\overline{H}^{b}\right)  \delta_{I}^{i}\delta_{j}%
^{J}\delta_{i}^{j}\right\vert ^{2} \nonumber\\
&  =\frac{1}{2}a\overline{H}H
\end{align*}
where
\begin{align*}
a  &  =\text{\textrm{tr}}\left(  k^{\ast\nu}k^{\nu}+k^{\ast e}k^{e}+3\left(
k^{\ast u}k^{u}+k^{\ast d}k^{d}\right)  \right) \\
c  &  =\text{\textrm{tr}}\left(  k^{\ast\nu_{R}}k^{\nu_{R}}\right)
\end{align*}
Next for the $a_{4}$ term, starting with the gauge kinetic energies we have
\begin{equation*}
g_{L}^{2}\left(  W_{\mu\nu L}^{\alpha}\right)  ^{2}+g_{R}^{2}\left(  W_{\mu\nu
R}^{\alpha}\right)  ^{2}+g^{2}\left(  V_{\mu\nu}^{m}\right)  ^{2}\rightarrow
g_{L}^{2}\left(  W_{\mu\nu L}^{\alpha}\right)  ^{2}+\frac{5}{3}g_{1}^{2}%
B_{\mu\nu}^{2}+g_{3}^{2}\left(  V_{\mu\nu}^{m}\right)  ^{2}%
\end{equation*}
where $m=1,\cdots,8$ for $V_{\mu\nu}^{m}$ restricted to the $SU(3)$ gauge
group. Next for the Higgs kinetic and quartic terms we have%

\begin{align*}
\ \nabla_{\mu}\Sigma_{aI}^{\overset{.}{c}K}\nabla^{\mu}\Sigma_{\overset{.}{c}%
K}^{aI}  &  \rightarrow a\nabla_{\mu}\overline{H}\nabla^{\mu}H\\
\frac{1}{2}\nabla_{\mu}H_{\overset{.}{a}I\overset{.}{b}J}\nabla^{\mu
}H^{\overset{.}{a}I\overset{.}{b}J}  &  \rightarrow\frac{1}{2}c\partial_{\mu
}\sigma\partial^{\mu}\sigma\\
\frac{1}{12}R\left(  H_{\overset{.}{a}I\overset{.}{c}K}H^{\overset{.}{c}%
K\overset{.}{a}I}+2\Sigma_{\overset{.}{a}I}^{cK}\Sigma_{cK}^{\overset{.}{a}%
I}\right)   &  \rightarrow\frac{1}{12}R\left(  2a\overline{H}H+c\sigma
^{2}\right) \\
\frac{1}{2}\left\vert H_{\overset{.}{a}I\overset{.}{c}K}H^{\overset{.}{c}%
K\overset{.}{b}J}\right\vert ^{2}  &  \rightarrow\frac{1}{2}d\sigma^{4}\\
2H_{\overset{.}{a}I\overset{.}{c}K}\Sigma_{bJ}^{\overset{.}{c}K}%
H^{\overset{.}{a}I\overset{.}{d}L}\Sigma_{\overset{.}{d}L}^{bJ}  &
\rightarrow2e\overline{H}H\sigma^{2}\\
\Sigma_{aI}^{\overset{.}{c}K}\Sigma_{\overset{.}{c}K}^{bJ}\Sigma
_{bJ}^{\overset{.}{dL}}\Sigma_{\overset{.}{d}L}^{aI}  &  \rightarrow b\left(
\overline{H}H\right).  ^{2}%
\end{align*}
Collecting all terms we end up with the bosonic action for the Standard
Model:
\begin{align*}
S_{\mathrm{b}}  &  =\frac{24}{\pi^{2}}F_{4}\Lambda^{4}%
{\displaystyle\int}
d^{4}x\sqrt{g}\nonumber\\
&  -\frac{2}{\pi^{2}}F_{2}\Lambda^{2}%
{\displaystyle\int}
d^{4}x\sqrt{g}\left(  R+\frac{1}{2}a\overline{H}H+\frac{1}{4}c\sigma
^{2}\right) \nonumber\\
&  +\frac{1}{2\pi^{2}}F_{0}%
{\displaystyle\int}
d^{4}x\sqrt{g}\left[  \frac{1}{30}\left(  -18C_{\mu\nu\rho\sigma}%
^{2}+11R^{\ast}R^{\ast}\right)  +\frac{5}{3}g_{1}^{2}B_{\mu\nu}^{2}+g_{2}%
^{2}\left(  W_{\mu\nu}^{\alpha}\right)  ^{2}+g_{3}^{2}\left(  V_{\mu\nu}%
^{m}\right)  ^{2}\right. \nonumber\\
&  \qquad\left.  +\frac{1}{6}aR\overline{H}H+b\left(  \overline{H}H\right)
^{2}+a\left\vert \nabla_{\mu}H_{a}\right\vert ^{2}+2e\overline{H}H\,\sigma
^{2}+\frac{1}{2}d\,\sigma^{4}+\frac{1}{12}cR\sigma^{2}+\frac{1}{2}c\left(
\partial_{\mu}\sigma\right)  ^{2}\right] \nonumber
\end{align*}
where%
\begin{align*}
b  &  =\text{\textrm{tr}}\left(  \left(  k^{\ast\nu}k^{\nu}\right)
^{2}+\left(  k^{\ast e}k^{e}\right)  ^{2}+3\left(  \left(  k^{\ast u}%
k^{u}\right)  ^{2}+\left(  k^{\ast d}k^{d}\right)  ^{2}\right)  \right) \\
d  &  =\text{\textrm{tr}}\left(  \left(  k^{\ast\nu_{R}}k^{\nu_{R}}\right)
^{2}\right) \\
e  &  =\text{\textrm{tr}}\left(  k^{\ast\nu}k^{\nu}k^{\ast\nu_{R}}k^{\nu_{R}%
}\right).
\end{align*}
This action completely agrees with the results in reference \cite{framework}.

\section{The potential and symmetry breaking}
\label{sect:pot}
We now study the resulting potential and try to investigate the possible
minima:
\begin{align*}
V &  =\frac{F_{0}}{2\pi^{2}}\left(  \frac{1}{2}\left\vert H_{\overset{.}{a}%
I\overset{.}{c}K}H^{\overset{.}{c}K\overset{.}{b}J}\right\vert ^{2}%
+2H_{\overset{.}{a}I\overset{.}{c}K}\Sigma_{bJ}^{\overset{.}{c}K}%
H^{\overset{.}{a}I\overset{.}{d}L}\Sigma_{\overset{.}{d}L}^{bJ}+\Sigma
_{aI}^{\overset{.}{c}K}\Sigma_{\overset{.}{c}K}^{bJ}\Sigma_{bJ}%
^{\overset{.}{d}L}\Sigma_{\overset{.}{d}L}^{aI}\right)  \nonumber\\
&  -\frac{F_{2}}{2\pi^{2}}\left(  H_{\overset{.}{a}I\overset{.}{c}%
K}H^{\overset{.}{c}K\overset{.}{a}I}+2\Sigma_{\overset{.}{a}I}^{cK}\Sigma
_{cK}^{\overset{.}{a}I}\right).
\end{align*}
However, the Higgs field here are not fundamental and we have to express the
potential in terms of the fundamental Higgs fields $\phi_{\overset{.}{a}}%
^{c},$ $\Delta_{\overset{.}{a}K}$ and $\Sigma_{K}^{I}.$ Expanding the
composite Higgs fields in terms of the fundamental ones, we have for the
quartic terms
\begin{equation*}
\frac{1}{2}\left\vert H_{\overset{.}{a}I\overset{.}{c}K}H^{\overset{.}{c}%
K\overset{.}{b}J}\right\vert ^{2}=\frac{1}{2}\left\vert k^{\nu_{R}}\right\vert
^{4}\left(  \Delta_{\overset{.}{a}K}\overline{\Delta}^{\overset{.}{a}L}%
\Delta_{\overset{.}{b}L}\overline{\Delta}^{\overset{.}{b}K}\right)  ^{2}%
\end{equation*}%
\begin{align*}
\Sigma_{aI}^{\overset{.}{c}K}\Sigma_{\overset{.}{c}K}^{bJ}\Sigma
_{bJ}^{\overset{.}{d}L}\Sigma_{\overset{.}{d}L}^{aI} &  =\left(  \left(
\left(  k^{\ast\nu}-k^{\ast u}\right)  \phi_{a}^{\overset{.}{c}}+\left(
k^{\ast e}-k^{\ast d}\right)  \widetilde{\phi}_{a}^{\overset{.}{c}}\right)
\Sigma_{I}^{K}+\left(  k^{\ast u}\phi_{a}^{\overset{.}{c}}+k^{\ast
d}\widetilde{\phi}_{a}^{\overset{.}{c}}\right)  \delta_{I}^{K}\right)
\nonumber\\
&  \left(  \left(  \left(  k^{\nu}-k^{u}\right)  \phi_{\overset{.}{c}}%
^{b}+\left(  k^{e}-k^{d}\right)  \widetilde{\phi}_{\overset{.}{c}}^{b}\right)
\Sigma_{K}^{J}+\left(  k^{u}\phi_{\overset{.}{c}}^{b}+k^{d}\widetilde{\phi
}_{\overset{.}{c}}^{b}\right)  \delta_{K}^{J}\right)  \nonumber\\
&  \left(  \left(  \left(  k^{\ast\nu}-k^{\ast u}\right)  \phi_{b}%
^{\overset{.}{d}}+\left(  k^{\ast e}-k^{\ast d}\right)  \widetilde{\phi}%
_{b}^{\overset{.}{d}}\right)  \Sigma_{J}^{L}+\left(  k^{\ast u}\phi
_{b}^{\overset{.}{d}}+k^{\ast d}\widetilde{\phi}_{b}^{\overset{.}{d}}\right)
\delta_{J}^{L}\right)  \nonumber\\
&  \left(  \left(  \left(  k^{\nu}-k^{u}\right)  \phi_{\overset{.}{d}}%
^{a}+\left(  k^{e}-k^{d}\right)  \widetilde{\phi}_{\overset{.}{d}}^{a}\right)
\Sigma_{L}^{I}+\left(  k^{u}\phi_{\overset{.}{d}}^{a}+k^{d}\widetilde{\phi
}_{\overset{.}{d}}^{a}\right)  \delta_{L}^{I}\right)
\end{align*}%
\begin{align*}
2H_{\overset{.}{a}I\overset{.}{c}K}\Sigma_{bJ}^{\overset{.}{c}K}%
H^{\overset{.}{a}I\overset{.}{d}L}\Sigma_{\overset{.}{d}L}^{bJ} &
=2\left\vert k^{\nu_{R}}\right\vert ^{2}\left(  \Delta_{\overset{.}{a}%
K}\overline{\Delta}^{\overset{.}{a}L}\Delta_{\overset{.}{c}I}\overline{\Delta
}^{\overset{.}{d}I}\right)  \nonumber\\
&  \left(  \left(  \left(  k^{\ast\nu}-k^{\ast u}\right)  \phi_{b}%
^{\overset{.}{c}}+\left(  k^{\ast e}-k^{\ast d}\right)  \widetilde{\phi}%
_{b}^{\overset{.}{c}}\right)  \Sigma_{J}^{K}+\left(  k^{\ast u}\phi
_{b}^{\overset{.}{c}}+k^{\ast d}\widetilde{\phi}_{b}^{\overset{.}{c}}\right)
\delta_{J}^{K}\right)  \nonumber\\
&  \left(  \left(  \left(  k^{\nu}-k^{u}\right)  \phi_{\overset{.}{d}}%
^{b}+\left(  k^{e}-k^{d}\right)  \widetilde{\phi}_{\overset{.}{d}}^{b}\right)
\Sigma_{L}^{J}+\left(  k^{u}\phi_{\overset{.}{d}}^{b}+k^{d}\widetilde{\phi
}_{\overset{.}{d}}^{b}\right)  \delta_{L}^{J}\right).
\end{align*}
Next we have the mass terms
\begin{equation*}
H_{\overset{.}{a}I\overset{.}{c}K}H^{\overset{.}{c}K\overset{.}{a}%
I}=\left\vert k^{\nu_{R}}\right\vert ^{2}\left(  \Delta_{\overset{.}{a}%
K}\overline{\Delta}^{\overset{.}{a}K}\right)  ^{2}%
\end{equation*}
and
\begin{align*}
2\Sigma_{\overset{.}{a}I}^{cK}\Sigma_{cK}^{\overset{.}{a}I}  & =2\left(
\left(  \left(  k^{\nu}-k^{u}\right)  \phi_{\overset{.}{a}}^{c}+\left(
k^{e}-k^{d}\right)  \widetilde{\phi}_{\overset{.}{a}}^{c}\right)  \Sigma
_{I}^{K}+\left(  k^{u}\phi_{\overset{.}{a}}^{c}+k^{d}\widetilde{\phi
}_{\overset{.}{a}}^{c}\right)  \delta_{I}^{K}\right)  \nonumber\\
& \left(  \left(  \left(  k^{\ast\nu}-k^{\ast u}\right)  \phi_{c}%
^{\overset{.}{a}}+\left(  k^{\ast e}-k^{\ast d}\right)  \widetilde{\phi}%
_{c}^{\overset{.}{a}}\right)  \Sigma_{K}^{I}+\left(  k^{\ast u}\phi
_{c}^{\overset{.}{a}}+k^{\ast d}\widetilde{\phi}_{c}^{\overset{.}{a}}\right)
\delta_{K}^{I}\right).
\end{align*}
The potential must be analyzed to determine all the possible minima that
breaks the symmetry $SU\left(  2\right)  _{R}\times SU\left(  2\right)
_{L}\times SU\left(  4\right)  .$ In this respect it is useful to determine
whether the symmetries of this model break correctly at high energies to the
Standard Model.

\begin{figure}[ptb]
\includegraphics[scale=.6]{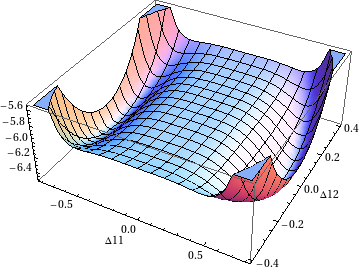}
\includegraphics[scale=.6]{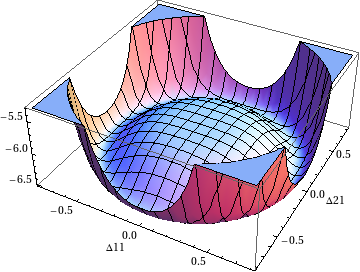}
\includegraphics[scale=.6]{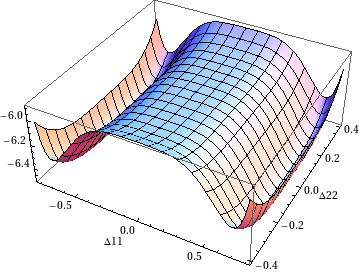}
\includegraphics[scale=.6]{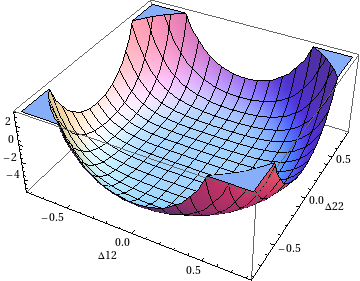}\caption{The
scalar potential in some of the $\Delta_{\dot a I}$-directions, with all other
fields at their SM-vevs as in Equation \eqref{eq:SM-vev}. We have put $k^{\nu
}= k^{e} = 1$ and $k^{\nu_{R}} = k^{u} = k^{d} = 2$. With these choices, the
Standard Model vacuum corresponds to $\Delta_{\dot1 1}=
\tfrac1 {\sqrt{2}}, \Sigma_1^1= 2, \phi_{\dot 1}^1 = \tfrac12$ and all other fields are zero. At this point the Hessian in the $\Delta$-directions is nonnegative. }%
\label{fig:potential-Delta}%
\end{figure}

Needless to say that it is difficult to determine all allowed vacua of this
potential, especially since there is dependence of order eight on the fields.
It is possible, however, to expand this potential around the vacuum that we
started with which breaks the gauge symmetry directly from $SU\left(
2\right)  _{R}\times SU\left(  2\right)  _{L}\times SU\left(  4\right)  $ to
$U\left(  1\right)  _{\mathrm{em}}\times SU\left(  3\right)  _{\mathrm{c}}$.
Explicitly, this vacuum is given by
\begin{align}
\label{eq:SM-vev} \left\langle \phi_{\overset{.}{a}}^{b}\right\rangle =v\delta_{\overset{.}{a}%
}^{\overset{.}{1}}\delta_{1}^{b}\qquad
\left\langle \Sigma_{J}^{I} \right\rangle =u\delta_{1}^{I}%
\delta_{J}^{1}\qquad
\left\langle \Delta_{\overset{.}{a}J} \right\rangle =w\delta_{\overset{.}{a}%
}^{\overset{.}{1}}\delta_{J}^{1}.
\end{align}
We have included several plots of the scalar potential in the $\Delta_{\dot a J}$-directions in Figure \ref{fig:potential-Delta}. A computation of the Hessian in the $\Delta$-directions shows that the SM-vev is indeed a local minimum.

The first order condition now arises as a vacuum solution of the spectral action as follows. We let the $\Delta$-fields take their vev according to the scalar potential, {\it i.e.} $\Delta_{\overset{.}{a}J} =w\delta_{\overset{.}{a}}^{\overset{.}{1}}\delta_{J}^{1}$.
Since $\Delta_{\dot a J}$ is in the $(2_R, 1_L,4)$ representation of $SU\left(
2\right)  _{R}\times SU\left(  2\right)  _{L}\times SU\left(  4\right)  $, this vacuum solution is only invariant under the subgroup
\begin{equation*}
\left \{ \left (  \left(\begin{smallmatrix} \lambda & 0 \\0 & \bar \lambda \end{smallmatrix} \right), u_L, \lambda \oplus \lambda^{-1/3} u \right)  : \lambda \in U(1), u_L \in SU(2), u \in SU(3) \right\}
\subset SU\left(
2\right)  _{R}\times SU\left(  2\right)  _{L}\times SU\left(  4\right) .
\end{equation*}
This is the spontaneous symmetry breaking to $U(1) \times SU(2)_L \times SU(3)_{\text c}$, thus selecting the subalgebra \eqref{subalg}. Note that unimodularity on $\mathcal U (\mathcal A)$ naturally induces unimodularity of the spectral Standard Model, hence it generates the correct hypercharges for the fermions.

\begin{figure}[ptb]
\includegraphics[scale=.6]{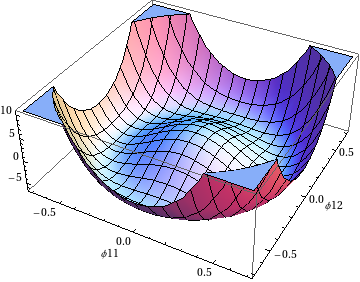}
\includegraphics[scale=.6]{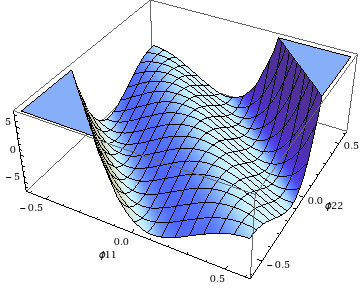}
\caption{The scalar
potential in the $\phi_{\dot a}^{b}$-directions, after the $\Sigma$ and $\Delta$-fields have acquired their SM-vevs as in Equation \eqref{eq:SM-vev}. Again, we have put $k^{\nu}= k^{e} = 1$ and $k^{\nu_{R}} = k^{u} = k^{d} = 2$. }%
\label{fig:potential-phi}%
\end{figure}

After the $\Delta$ and $\Sigma$-fields have acquired their vevs, there is a remaining scalar potential for the $\phi$-fields, which is depicted in Figure \ref{fig:potential-phi}. As with the Standard Model Higgs sector, the selection of a minimum further breaks the symmetry from $U(1) \times SU(2)_L \times SU(3)_{\text{c}}$ to $U(1)_{\text{em}} \times SU(3)_{\text{c}}$. The plot on the right in Figure \ref{fig:potential-phi} suggests that, instead of the SM-vacuum, the vevs of the $\phi$-fields can also be taken of the form
 \begin{equation*}
\left\langle \phi_{\overset{.}{a}}^{b}\right\rangle =v\delta_{\overset{.}{a}%
}^{\overset{.}{1}}\delta_{1}^{b}+v^{\prime}\delta_{\overset{.}{a}%
}^{\overset{.}{2}}\delta_{2}^{b}.
\end{equation*}

Let us see which of the gauge fields acquire non-zero mass after spontaneous symmetry breaking, by expanding around the Standard Model vacuum
\begin{align*}
\phi_{\overset{.}{a}}^{b}  &  =v\delta_{\overset{.}{a}}^{\overset{.}{1}}%
\delta_{1}^{b}+H_{\overset{.}{a}}^{b}\\
\Sigma_{J}^{I}  &  =u\delta_{1}^{I}\delta_{J}^{1}+M_{I}^{J}\\
\Delta_{\overset{.}{a}J}  &  =w\delta_{\overset{.}{a}}^{\overset{.}{1}}%
\delta_{J}^{1}+N_{\overset{.}{a}J}%
\end{align*}
and keep only terms of up to order $4.$ First we look at the kinetic term%
\begin{align*}
\nabla_{\mu}H_{\overset{.}{a}I\overset{.}{b}J}  &  =\partial_{\mu
}H_{\overset{.}{a}I\overset{.}{b}J}-\frac{i}{2}g_{R}W_{\mu R}^{\alpha}\left(
\sigma^{\alpha}\right)  _{\overset{.}{a}}^{\overset{.}{c}}H_{\overset{.}{c}%
I\overset{.}{b}J}-\frac{i}{2}g_{R}W_{\mu R}^{\alpha}\left(  \sigma^{\alpha
}\right)  _{\overset{.}{b}}^{\overset{.}{c}}H_{\overset{.}{a}I\overset{.}{c}%
J}\nonumber\\
&  -\frac{i}{2}gV_{\mu}^{m}\left(  \lambda^{m}\right)  _{I}^{K}%
H_{\overset{.}{a}K\overset{.}{b}J}-\frac{i}{2}gV_{\mu}^{m}\left(  \lambda
^{m}\right)  _{J}^{K}H_{\overset{.}{a}I\overset{.}{b}K}.%
\end{align*}
To lowest orders we have
\begin{align*}
H_{\overset{.}{a}I\overset{.}{b}J}  &  =\left(  k^{\ast\nu_{R}}\right)
^{2}\left(  w\delta_{\overset{.}{a}}^{\overset{.}{1}}\delta_{J}^{1}%
+N_{\overset{.}{a}J}\right)  \left(  w\delta_{\overset{.}{b}}^{\overset{.}{1}%
}\delta_{I}^{1}+N_{\overset{.}{b}I}\right) \\
&  =\left(  k^{\ast\nu_{R}}\right)  ^{2}\left(  w^{2}\delta_{\overset{.}{a}%
}^{\overset{.}{1}}\delta_{J}^{1}\delta_{\overset{.}{b}}^{\overset{.}{1}}%
\delta_{I}^{1}+w\delta_{\overset{.}{a}}^{\overset{.}{1}}\delta_{J}%
^{1}N_{\overset{.}{b}I}+w\delta_{\overset{.}{b}}^{\overset{.}{1}}\delta
_{I}^{1}N_{\overset{.}{a}J}+N_{\overset{.}{a}J}N_{\overset{.}{b}I}\right)
\nonumber
\end{align*}
and so
\begin{align*}
\nabla_{\mu}H_{\overset{.}{a}I\overset{.}{b}J}  &  =\left(  k^{\ast\nu_{R}%
}\right)  ^{2}w\left(  \ \delta_{\overset{.}{a}}^{\overset{.}{1}}\delta
_{J}^{1}\partial_{\mu}N_{\overset{.}{b}I}+\ \delta_{\overset{.}{b}%
}^{\overset{.}{1}}\delta_{I}^{1}\partial_{\mu}N_{\overset{.}{a}J}-\frac{i}%
{2}g_{R}W_{\mu R}^{\alpha}\left(  \sigma^{\alpha}\right)  _{\overset{.}{a}%
}^{\overset{.}{1}}w\delta_{J}^{1}\delta_{\overset{.}{b}}^{\overset{.}{1}%
}\delta_{I}^{1}-\frac{i}{2}g_{R}W_{\mu R}^{\alpha}\left(  \sigma^{\alpha
}\right)  _{\overset{.}{b}}^{\overset{.}{1}}w\delta_{\overset{.}{a}%
}^{\overset{.}{1}}\delta_{J}^{1}\delta_{I}^{1}\right. \nonumber\\
&  \qquad\left.  -\frac{i}{2}gV_{\mu}^{m}\left(  \lambda^{m}\right)  _{I}%
^{1}w\delta_{\overset{.}{a}}^{\overset{.}{1}}\delta_{J}^{1}\delta
_{\overset{.}{b}}^{\overset{.}{1}}-\frac{i}{2}gV_{\mu}^{m}\left(  \lambda
^{m}\right)  _{J}^{1}w\delta_{\overset{.}{a}}^{\overset{.}{1}}\delta
_{\overset{.}{b}}^{\overset{.}{1}}\delta_{I}^{1}\right) \nonumber\\
&  =\left(  k^{\ast\nu_{R}}\right)  ^{2}w\left(  2\ \delta_{\overset{.}{a}%
}^{\overset{.}{1}}\delta_{J}^{1}\delta_{\overset{.}{b}}^{\overset{.}{1}}%
\delta_{I}^{1}\partial_{\mu}N_{\overset{.}{1}1}+\delta_{\overset{.}{a}%
}^{\overset{.}{1}}\delta_{J}^{1}\delta_{\overset{.}{b}}^{\overset{.}{2}}%
\delta_{I}^{1}\partial_{\mu}N_{\overset{.}{2}1}+\ \delta_{\overset{.}{a}%
}^{\overset{.}{1}}\delta_{J}^{1}\delta_{I}^{i}\delta_{\overset{.}{b}%
}^{\overset{.}{1}}\partial_{\mu}N_{\overset{.}{1}i}+\delta_{\overset{.}{a}%
}^{\overset{.}{1}}\delta_{J}^{1}\delta_{I}^{i}\delta_{\overset{.}{b}%
}^{\overset{.}{2}}\partial_{\mu}N_{\overset{.}{2}i}\right. \nonumber\\
&  +\ \delta_{\overset{.}{b}}^{\overset{.}{1}}\delta_{I}^{1}\ \delta
_{\overset{.}{a}}^{\overset{.}{1}}\delta_{J}^{j}\partial_{\mu}%
N_{\overset{.}{1}j}+\delta_{\overset{.}{b}}^{\overset{.}{1}}\delta_{I}%
^{1}\ \delta_{\overset{.}{a}}^{\overset{.}{2}}\delta_{J}^{1}\partial_{\mu
}N_{\overset{.}{2}1}+\delta_{\overset{.}{b}}^{\overset{.}{1}}\delta_{I}%
^{1}\ \delta_{\overset{.}{a}}^{\overset{.}{2}}\delta_{J}^{j}\partial_{\mu
}N_{\overset{.}{2}j}-ig_{R}W_{\mu R}^{3}w\delta_{\overset{.}{a}}%
^{\overset{.}{1}}\delta_{J}^{1}\delta_{\overset{.}{b}}^{\overset{.}{1}}%
\delta_{I}^{1}\nonumber\\
&  -\frac{i}{2}g_{R}W_{\mu R}^{-}w\delta_{\overset{.}{a}}^{\overset{.}{2}%
}\delta_{J}^{1}\delta_{\overset{.}{b}}^{\overset{.}{1}}\delta_{I}^{1}-\frac
{i}{2}g_{R}W_{\mu R}^{-}\delta_{\overset{.}{b}}^{\overset{.}{2}}%
w\delta_{\overset{.}{a}}^{\overset{.}{1}}\delta_{J}^{1}\delta_{I}%
^{1}\nonumber\\
&  -i\left(  g_{R}W_{\mu R}^{3}+g\sqrt{\frac{3}{2}}V_{\mu}^{15}\right)
\delta_{\overset{.}{a}}^{\overset{.}{1}}\delta_{J}^{1}\delta_{\overset{.}{b}%
}^{\overset{.}{1}}\delta_{I}^{1}-\frac{i}{2}gV_{\mu}^{m}\left(  \lambda
^{m}\right)  _{i}^{1}w\delta_{\overset{.}{a}}^{\overset{.}{1}}\delta_{J}%
^{1}\delta_{\overset{.}{b}}^{\overset{.}{1}}\delta_{I}^{i}%
\end{align*}
from which it is clear that if we write
\begin{align*}
g_{R}W_{\mu R}^{3}  &  =g_{1}B_{\mu}+g_{1}^{\prime}Z_{\mu}^{\prime}\\
g\sqrt{\frac{3}{2}}V_{\mu}^{15}  &  =-g_{1}B_{\mu}+g_{1}^{\prime}Z_{\mu
}^{\prime}%
\end{align*}
then the vector $B_{\mu}$ will not get a mass term while the fields $W_{\mu
R}^{\pm},$ $Z_{\mu}^{\prime},$ $V_{\mu}^{m}\left(  \lambda^{m}\right)
_{i}^{1}$ (these are the fields in the coset of $\frac{SU\left(  2\right)
_{R}\times SU\left(  4\right)  }{SU\left(  3\right)  \times U\left(  1\right)
}$) will all become massive, with mass of order $w^{2}$ as can be seen from
the kinetic term
\begin{align*}
\nabla_{\mu}\Sigma_{\overset{.}{a}I}^{bJ}  &  =\partial_{\mu}\Sigma
_{\overset{.}{a}I}^{bJ}-\frac{i}{2}g_{R}W_{\mu R}^{\alpha}\left(
\sigma^{\alpha}\right)  _{\overset{.}{a}}^{\overset{.}{c}}\Sigma
_{\overset{.}{c}I}^{bJ}+\frac{i}{2}g_{R}W_{\mu R}^{\alpha}\left(
\sigma^{\alpha}\right)  _{c}^{b}\Sigma_{\overset{.}{a}I}^{cJ}\nonumber\\
&  -\frac{i}{2}gV_{\mu}^{m}\left(  \lambda^{m}\right)  _{I}^{K}\Sigma
_{\overset{.}{a}K}^{bJ}+\frac{i}{2}gV_{\mu}^{m}\left(  \lambda^{m}\right)
_{K}^{J}\Sigma_{\overset{.}{a}I}^{bK}.%
\end{align*}
To lowest orders we have%
\begin{align*}
\Sigma_{\overset{.}{a}I}^{bJ}  &  =\left(  \left(  \left(  k^{\nu}%
-k^{u}\right)  \phi_{\overset{.}{a}}^{b}+\left(  k^{e}-k^{d}\right)
\widetilde{\phi}_{\overset{.}{a}}^{b}\right)  \Sigma_{I}^{J}+\left(  k^{u}%
\phi_{\overset{.}{a}}^{b}+k^{d}\widetilde{\phi}_{\overset{.}{a}}^{b}\right)
\delta_{I}^{J}\right) \nonumber\\
&  =\left(  \left(  k^{\nu}-k^{u}\right)  \left(  v\delta_{\overset{.}{a}%
}^{\overset{.}{1}}\delta_{1}^{b}+H_{\overset{.}{a}}^{b}\right)  +\left(
k^{e}-k^{d}\right)  \left(  v\delta_{\overset{.}{a}}^{\overset{.}{2}}%
\delta_{2}^{b}+\widetilde{H}_{\overset{.}{a}}^{b}\right)  \right)  \left(
u\delta_{1}^{J}\delta_{I}^{1}+M_{I}^{J}\right) \nonumber\\
&  +\left(  k^{u}\left(  v\delta_{\overset{.}{a}}^{\overset{.}{1}}\delta
_{1}^{b}+H_{\overset{.}{a}}^{b}\right)  +k^{d}\left(  v\delta_{\overset{.}{a}%
}^{\overset{.}{2}}\delta_{2}^{b}+\widetilde{H}_{\overset{.}{a}}^{b}\right)
\right)  \delta_{I}^{J}\nonumber\\
&  =v\left(  \left(  \left(  k^{\nu}-k^{u}\right)  \delta_{\overset{.}{a}%
}^{\overset{.}{1}}\delta_{1}^{b}+\left(  k^{e}-k^{d}\right)  \delta
_{\overset{.}{a}}^{\overset{.}{2}}\delta_{2}^{b}\right)  u\delta_{1}^{J}%
\delta_{I}^{1}+\left(  k^{u}\delta_{\overset{.}{a}}^{\overset{.}{1}}\delta
_{1}^{b}+k^{d}\delta_{\overset{.}{a}}^{\overset{.}{2}}\delta_{2}^{b}\right)
\delta_{I}^{J}\right) \nonumber\\
&  +\left(  \left(  k^{\nu}-k^{u}\right)  H_{\overset{.}{a}}^{b}+\left(
k^{e}-k^{d}\right)  \widetilde{H}_{\overset{.}{a}}^{b}\right)  u\delta_{1}%
^{J}\delta_{I}^{1}+\left(  k^{u}H_{\overset{.}{a}}^{b}+k^{d}\widetilde{H}%
_{\overset{.}{a}}^{b}\right)  \delta_{I}^{J}\nonumber\\
&  +v\left(  \left(  k^{\nu}-k^{u}\right)  \delta_{\overset{.}{a}%
}^{\overset{.}{1}}\delta_{1}^{b}+\left(  k^{e}-k^{d}\right)  \delta
_{\overset{.}{a}}^{\overset{.}{2}}\delta_{2}^{b}\right)  M_{I}^{J}%
\end{align*}%
\begin{align*}
\nabla_{\mu}\Sigma_{\overset{.}{a}I}^{bJ}  &  =\left(  \left(  k^{\nu}%
-k^{u}\right)  \partial_{\mu}H_{\overset{.}{a}}^{b}+\left(  k^{e}%
-k^{d}\right)  \partial_{\mu}\widetilde{H}_{\overset{.}{a}}^{b}\right)
u\delta_{1}^{J}\delta_{I}^{1}+\left(  k^{u}\partial_{\mu}H_{\overset{.}{a}%
}^{b}+k^{d}\partial_{\mu}\widetilde{H}_{\overset{.}{a}}^{b}\right)  \delta
_{I}^{J}\nonumber\\
&  +v\left(  \left(  k^{\nu}-k^{u}\right)  \delta_{\overset{.}{a}%
}^{\overset{.}{1}}\delta_{1}^{b}+\left(  k^{e}-k^{d}\right)  \delta
_{\overset{.}{a}}^{\overset{.}{2}}\delta_{2}^{b}\right)  \partial_{\mu}%
M_{I}^{J}\nonumber\\
&  -\frac{i}{2}vg_{R}W_{\mu R}^{\alpha}\left(  \sigma^{\alpha}\right)
_{\overset{.}{a}}^{\overset{.}{c}}\left(  \left(  \left(  k^{\nu}%
-k^{u}\right)  \delta_{\overset{.}{c}}^{\overset{.}{1}}\delta_{1}^{b}+\left(
k^{e}-k^{d}\right)  \delta_{\overset{.}{c}}^{\overset{.}{2}}\delta_{2}%
^{b}\right)  u\delta_{1}^{J}\delta_{I}^{1}+\left(  k^{u}\delta_{\overset{.}{c}%
}^{\overset{.}{1}}\delta_{1}^{b}+k^{d}\delta_{\overset{.}{c}}^{\overset{.}{2}%
}\delta_{2}^{b}\right)  \delta_{I}^{J}\right) \nonumber\\
&  +\frac{i}{2}vg_{L}W_{\mu L}^{\alpha}\left(  \sigma^{\alpha}\right)
_{c}^{b}\left(  \left(  \left(  k^{\nu}-k^{u}\right)  \delta_{\overset{.}{a}%
}^{\overset{.}{1}}\delta_{1}^{c}+\left(  k^{e}-k^{d}\right)  \delta
_{\overset{.}{a}}^{\overset{.}{2}}\delta_{2}^{c}\right)  u\delta_{1}^{J}%
\delta_{I}^{1}+\left(  k^{u}\delta_{\overset{.}{a}}^{\overset{.}{1}}\delta
_{1}^{c}+k^{d}\delta_{\overset{.}{a}}^{\overset{.}{2}}\delta_{2}^{c}\right)
\delta_{I}^{J}\right) \nonumber\\
&  -\frac{i}{2}vgV_{\mu}^{m}\left(  \lambda^{m}\right)  _{I}^{K}\left(
\left(  \left(  k^{\nu}-k^{u}\right)  \delta_{\overset{.}{a}}^{\overset{.}{1}%
}\delta_{1}^{b}+\left(  k^{e}-k^{d}\right)  \delta_{\overset{.}{a}%
}^{\overset{.}{2}}\delta_{2}^{b}\right)  u\delta_{1}^{J}\delta_{K}^{1}+\left(
k^{u}\delta_{\overset{.}{a}}^{\overset{.}{1}}\delta_{1}^{b}+k^{d}%
\delta_{\overset{.}{a}}^{\overset{.}{2}}\delta_{2}^{b}\right)  \delta_{K}%
^{J}\right) \nonumber\\
&  +\frac{i}{2}gvV_{\mu}^{m}\left(  \lambda^{m}\right)  _{K}^{J}\left(
\left(  \left(  k^{\nu}-k^{u}\right)  \delta_{\overset{.}{a}}^{\overset{.}{1}%
}\delta_{1}^{b}+\left(  k^{e}-k^{d}\right)  \delta_{\overset{.}{a}%
}^{\overset{.}{2}}\delta_{2}^{b}\right)  u\delta_{1}^{K}\delta_{I}^{1}+\left(
k^{u}\delta_{\overset{.}{a}}^{\overset{.}{1}}\delta_{1}^{b}+k^{d}%
\delta_{\overset{.}{a}}^{\overset{.}{2}}\delta_{2}^{b}\right)  \delta_{I}%
^{K}\right).
\end{align*}
For simplicity we will set $u=1.$ Isolating the gauge dependent part%
\begin{align*}
\nabla_{\mu}\Sigma_{\overset{.}{1}1}^{11}  &  \supset-\frac{i}{2}v\left(
g_{R}W_{\mu R}^{3}-g_{L}W_{\mu L}^{3}\right)  k^{\nu}\\
\nabla_{\mu}\Sigma_{\overset{.}{2}1}^{21}  &  \supset\frac{i}{2}v\left(
g_{R}W_{\mu R}^{3}-g_{L}W_{\mu L}^{3}\right)  k^{e}\\
\nabla_{\mu}\Sigma_{\overset{.}{1}i}^{11}  &  \supset-\frac{i}{2}vgV_{\mu}%
^{m}\left(  \lambda^{m}\right)  _{i}^{1}\left(  k^{\nu}-k^{u}\right) \\
\nabla_{\mu}\Sigma_{\overset{.}{1}i}^{1j}  &  \supset-\frac{i}{2}v\left(
g_{R}W_{\mu R}^{3}-g_{L}W_{\mu L}^{3}\right)  k^{u}\delta_{i}^{j}\\
\nabla_{\mu}\Sigma_{\overset{.}{2}i}^{2j}  &  \supset\frac{i}{2}v\left(
g_{R}W_{\mu R}^{3}-g_{L}W_{\mu L}^{3}\right)  k^{d}\delta_{i}^{j}\\
\nabla_{\mu}\Sigma_{\overset{.}{1}1}^{21}  &  \supset-\frac{i}{2}v\left(
g_{R}W_{\mu R}^{-}-g_{L}W_{\mu L}^{-}\right)  k^{\nu}\\
\nabla_{\mu}\Sigma_{\overset{.}{2}1}^{11}  &  \supset-\frac{i}{2}v\left(
g_{R}W_{\mu R}^{+}-g_{L}W_{\mu L}^{+}\right)  k^{e}\\
\nabla_{\mu}\Sigma_{\overset{.}{2}i}^{11}  &  \supset0.
\end{align*}
Noticing that $g_{R}W_{\mu R}^{3}-g_{L}W_{\mu L}^{3}=\left(  g_{1}B_{\mu
}-g_{L}W_{\mu L}^{3}\right)  +g_{1}^{\prime}Z_{\mu}^{\prime}$ shows that the
$Z_{\mu}$ vector gets a mass of order of the weak scale $gv$ while the $W_{\mu
R}^{\pm}$ and $Z_{\mu}^{\prime}$ will get a small correction to its mass of
order $gw.$ Thus we get the correct gauge breaking pattern with the gauge
fields $W_{\mu L}$ and $Z$ of the Standard model having masses of the order of
the electroweak scale. It is important, however, to see explicitly that the
mixing between the $Z$ and $Z^{\prime}$ vectors and $W_{L}^{\pm}$, $W_{R}%
^{\pm}$ are suppressed.

It remains to minimize the potential to determine all possible minima as well
as studying the unified model and check whether it allows for unification of
coupling constants%
\[
g_{R}=g_{L}=g
\]
in addition to determining the top quark mass and Higgs mass. Obviously, this
model deserves careful analysis, which will be the subject of future work.

We conclude that the study of noncommutative spaces based on a product of a
continuous four dimensional manifold times a finite space of $KO$-dimension
$6,$ without the first order condition gives rise to almost unique possibility
in the form of a Pati-Salam type model. This provides a setting for
unification avoiding the desert and which goes beyond the SM. In addition one
of the vacua of the Higgs fields gives rise at low energies to a Dirac
operator satisfying the first order condition. In this way, the first order
condition arises as a spontaneously broken phase of higher symmetry and is not
imposed from outside.

\section{Appendix: Detailed calculations for the practitioner}
\label{sect:app}
For the benefit of the reader, we shall present in this appendix a detailed
derivation of the Dirac operator and the spectral action for the
noncommutative space on $\mathbb{H}_{R}\oplus\mathbb{H}_{L}\oplus M_{4}\left(
\mathbb{C}\right)  .$

For $A_{\left(  1\right)  }$ we have the definition
\begin{equation}
\left(  A_{\left(  1\right)  }\right)  _{M}^{\quad N\quad}=%
{\displaystyle\sum}
a_{M}^{P}\left[  D,b\right]  _{P}^{N}%
\end{equation}
where
\begin{equation}
a_{M}^{N}=\left(
\begin{array}
[c]{cc}%
X_{\alpha}^{\prime\beta}\delta_{I}^{J} & 0\\
0 & \delta_{\alpha^{\prime}}^{\beta^{\prime}}Y_{I^{\prime}}^{\prime J^{\prime
}}%
\end{array}
\right)
\end{equation}
which in terms of components give%
\begin{align}
\left(  A_{\left(  1\right)  }\right)  _{\alpha I}^{\beta J}  &  =%
{\displaystyle\sum}
a_{\alpha I}^{\gamma K}\left(  D_{\gamma K}^{\delta L}b_{\delta L}^{\beta
J}-b_{\gamma K}^{\delta L}D_{\delta L}^{\beta J}\right) \nonumber\\
&  =%
{\displaystyle\sum}
X_{\alpha}^{^{\prime}\gamma}\left(  D_{\gamma I}^{\delta J}X_{\delta}^{\beta
}-X_{\gamma}^{\delta}D_{\delta I}^{\beta J}\right)
\end{align}
where we\ use the notation for $b$ to be the same as that of $a$ without
primes (i.e. $X^{\prime}\rightarrow X,$ $Y^{\prime}\rightarrow Y$). Since
$D_{\alpha I}^{\beta J}$ is non vanishing when connecting a dotted index
$\overset{.}{a}$ to $a,$ (cf. \eqref{eq:DFpp}) we have the non-vanishing components%
\begin{align}
\left(  A_{\left(  1\right)  }\right)  _{\overset{.}{a}I}^{bJ}  &  =%
{\displaystyle\sum}
X_{\overset{.}{a}}^{^{\prime}\overset{.}{c}}\left(  D_{\overset{.}{c}I}%
^{dJ}X_{d}^{b}-X_{\overset{.}{c}}^{\overset{.}{d}}D_{\overset{.}{d}I}%
^{bJ}\right) \nonumber\\
&  =\delta_{I}^{1}\delta_{1}^{J}\left(
{\displaystyle\sum}
X_{\overset{.}{a}}^{^{\prime}\overset{.}{c}}\left(  \left(  \delta
_{\overset{.}{a}}^{\overset{.}{1}}\delta_{1}^{d}k^{\nu}+\delta_{\overset{.}%
{a}}^{\overset{.}{2}}\delta_{2}^{d}k^{e}\right)  X_{d}^{b}\right)
-X_{\overset{.}{c}}^{\overset{.}{d}}\left(  \delta_{\overset{.}{d}}%
^{\overset{.}{1}}\delta_{1}^{b}k^{\nu}+\delta_{\overset{.}{d}}^{\overset{.}%
{2}}\delta_{2}^{b}k^{e}\right)  \right) \nonumber\\
&  +\delta_{I}^{i}\delta_{j}^{J}\delta_{i}^{j}\left(
{\displaystyle\sum}
X_{\overset{.}{a}}^{^{\prime}\overset{.}{c}}\left(  \left(  \delta
_{\overset{.}{a}}^{\overset{.}{1}}\delta_{1}^{b}k^{u}+\delta_{\overset{.}{a}%
}^{\overset{.}{2}}\delta_{2}^{b}k^{d}\right)  X_{d}^{b}\right)  -X_{\overset
{.}{c}}^{\overset{.}{d}}\left(  \delta_{\overset{.}{d}}^{\overset{.}{1}}%
\delta_{1}^{b}k^{u}+\delta_{\overset{.}{d}}^{\overset{.}{2}}\delta_{2}%
^{b}k^{d}\right)  \right) \nonumber\\
&  =\delta_{I}^{1}\delta_{1}^{J}\left(  k^{\nu}\phi_{\overset{.}{a}}^{b}%
+k^{e}\widetilde{\phi}_{\overset{.}{a}}^{b}\right)  +\delta_{I}^{i}\delta
_{j}^{J}\delta_{i}^{j}\left(  k^{u}\phi_{\overset{.}{a}}^{b}+k^{d}%
\widetilde{\phi}_{\overset{.}{a}}^{b}\right)
\end{align}
where
\begin{align}
\phi_{\overset{.}{a}}^{b}  &  =%
{\displaystyle\sum}
X_{\overset{.}{a}}^{^{\prime}\overset{.}{1}}X_{1}^{b}-X_{\overset{.}{a}%
}^{^{\prime}\overset{.}{c}}X_{\overset{.}{c}}^{\overset{.}{1}}\delta_{1}^{b}\\
\widetilde{\phi}_{\overset{.}{a}}^{b}  &  =%
{\displaystyle\sum}
X_{\overset{.}{a}}^{^{\prime}\overset{.}{2}}X_{2}^{b}-X_{\overset{.}{a}%
}^{^{\prime}\overset{.}{c}}X_{\overset{.}{c}}^{\overset{.}{2}}\delta_{2}^{b}%
\end{align}
We can check that
\begin{equation}
\widetilde{\phi}_{\overset{.}{a}}^{b}=\sigma_{2}\overline{\phi}_{\overset{.}{a}%
}^{b}\sigma_{2}%
\end{equation}
For example
\begin{align}
\widetilde{\phi}_{\overset{.}{1}}^{1}  &  =%
{\displaystyle\sum}
X_{\overset{.}{1}}^{^{\prime}\overset{.}{2}}X_{2}^{1}\nonumber\\
&  =%
{\displaystyle\sum}
\overline{X}_{\overset{.}{2}}^{^{\prime}\overset{.}{1}}\overline{X}_{1}%
^{2}\nonumber\\
&  =\overline{\phi}_{\overset{.}{2}}^{2}%
\end{align}
using the quaternionic property of the $X.$ Note that $\phi_{\overset{.}{a}%
}^{b}$ is in the $\left(  2_{R},2_{L},1\right)  $ representation of $SU\left(
2\right)  _{R}\times SU\left(  2\right)  _{L}\times SU\left(  4\right)  $.

Similarly we have
\begin{equation}
\left(A_{(1)}\right)_{aI}^{\overset{.}{b}J}=\left(  \left(A_{(1)}\right)_{\overset{.}{b}J}^{aI}\right)  ^{\ast}%
\end{equation}
(In reality one obtains an expression for $\left(A_{(1)}\right)_{aI}^{\overset{.}{b}J}$ in terms
of $\phi_{\overset{.}{a}}^{\prime b}$ which is expressed in terms of the $X,$
but the hermiticity of the Dirac operator forces the above relation and
imposes a constraint on the $X.$ )

Next we have using \eqref{eq:DFppbar}
\begin{align}
\left(  A_{\left(  1\right)  }\right)  _{\alpha I}^{\beta^{\prime}J^{\prime}}
&  =%
{\displaystyle\sum}
a_{\alpha I}^{\gamma K}\left(  D_{\gamma K}^{\delta^{\prime}L^{\prime}%
}b_{\delta^{\prime}L^{\prime}}^{\beta^{\prime}J^{\prime}}-b_{\gamma K}^{\delta
L}D_{\delta L}^{\beta^{\prime}J\prime}\right) \nonumber\\
&  =%
{\displaystyle\sum}
X_{\alpha}^{\prime\gamma}\left(  D_{\gamma I}^{\beta^{\prime}L^{\prime}%
}Y_{L^{\prime}}^{J^{\prime}}-X_{\gamma}^{\delta}D_{\delta I}^{\beta^{\prime
}J\prime}\right) \nonumber\\
&  =k^{\ast\nu_{R}}%
{\displaystyle\sum}
X_{\alpha}^{\prime\gamma}\left(  \left(  \delta_{\gamma}^{\overset{.}{1}%
}\delta_{\overset{.}{1}^{\prime}}^{\beta^{\prime}}\delta_{I}^{1}%
\delta_{1^{\prime}}^{L^{\prime}}\right)  Y_{L^{\prime}}^{J^{\prime}}%
-X_{\gamma}^{\delta}\left(  \delta_{\delta}^{\overset{.}{1}}\delta
_{\overset{.}{1}^{\prime}}^{\beta^{\prime}}\delta_{I}^{1}\delta_{1^{\prime}%
}^{J^{\prime}}\right)  \right) \nonumber\\
&  =k^{\ast\nu_{R}}\delta_{\alpha}^{\overset{.}{a}}\delta_{\overset{.}%
{1}^{\prime}}^{\beta^{\prime}}\delta_{I}^{1}%
{\displaystyle\sum}
\left(  X_{\overset{.}{a}}^{\prime\overset{.}{1}}Y_{1^{\prime}}^{J^{\prime}%
}-X_{\overset{.}{a}}^{\prime\overset{.}{c}}X_{\overset{.}{c}}^{\overset{.}{1}%
}\delta_{1^{\prime}}^{J^{\prime}}\right)
\end{align}%
\begin{equation}
\left(  A_{\left(  1\right)  }\right)  _{\overset{.}{a}I}^{\overset{.}%
{b}^{\prime}J^{\prime}}=k^{\ast\nu_{R}}\delta_{\overset{.}{1}^{\prime}%
}^{\overset{.}{b}^{\prime}}\delta_{I}^{1}\Delta_{\overset{.}{a}}^{\quad
J^{\prime}}%
\end{equation}
where
\begin{equation}
\Delta_{\overset{.}{a}}^{\quad J^{\prime}}=%
{\displaystyle\sum}
\left(  X_{\overset{.}{a}}^{\prime\overset{.}{1}}Y_{1^{\prime}}^{J^{\prime}%
}-X_{\overset{.}{a}}^{\prime\overset{.}{c}}X_{\overset{.}{c}}^{\overset{.}{1}%
}\delta_{1^{\prime}}^{J^{\prime}}\right)  \equiv\Delta_{\overset{.}{a}J}%
\end{equation}
which is in the $\left(  2_{R},1_{L},4\right)  $ representation of $SU\left(
2\right)  _{R}\times SU\left(  2\right)  _{L}\times SU\left(  4\right)  .$
Again, we can compute $\left(A_{(1)}\right)_{\alpha^{\prime}I^{\prime}}^{\beta J}$, which gives a
similar expression, but using hermiticity we write
\begin{equation}
\left(A_{(1)}\right)_{\alpha^{\prime}I^{\prime}}^{\beta J}=\left(  \left(A_{(1)}\right)_{\beta J}^{\alpha^{\prime
}I^{\prime}}\right)  ^{\ast}=k^{\nu_{R}}\delta_{\overset{.}{b}}^{\beta}%
\delta_{\alpha^{\prime}}^{1^{\prime}}\delta_{1}^{J}\Delta_{\quad I^{\prime}%
}^{\overset{.}{b}}%
\end{equation}
In the conjugate space we have
\begin{align}
\left(  A_{\left(  1\right)  }\right)  _{\alpha^{\prime}I^{\prime}}%
^{\beta^{\prime}J^{\prime}}  &  =%
{\displaystyle\sum}
a_{\alpha^{\prime}I^{\prime}}^{\gamma^{\prime}K^{\prime}}\left(
D_{\gamma^{\prime}K^{\prime}}^{\delta^{\prime}L^{\prime}}b_{\delta^{\prime
}L^{\prime}}^{\beta^{\prime}J^{\prime}}-b_{\gamma^{\prime}K^{\prime}}%
^{\delta^{\prime}L^{\prime}}D_{\delta^{\prime}L^{\prime}}^{\beta^{\prime
}J^{\prime}}\right) \nonumber\\
&  =%
{\displaystyle\sum}
Y_{I^{\prime}}^{\prime K^{\prime}}\left(  D_{\alpha^{\prime}K^{\prime}}%
^{\beta^{\prime}L^{\prime}}Y_{L^{\prime}}^{J^{\prime}}-Y_{K^{\prime}%
}^{L^{\prime}}D_{\alpha^{\prime}L^{\prime}}^{\beta^{\prime}J^{\prime}}\right)
\end{align}
The only non-vanishing expression would involve a $D$ with mixed $a^{\prime}$
and $\overset{.}{b}^{\prime}$%
\begin{align}
\left(  A_{\left(  1\right)  }\right)  _{\overset{.}{a}^{\prime}I^{\prime}%
}^{b^{\prime}J^{\prime}}  &  =%
{\displaystyle\sum}
Y_{I^{\prime}}^{\prime K^{\prime}}\left(  D_{\overset{.}{a}^{\prime}K^{\prime
}}^{b^{\prime}L^{\prime}}Y_{L^{\prime}}^{J^{\prime}}-Y_{K^{\prime}}%
^{L^{\prime}}D_{\overset{.}{a}^{\prime}L^{\prime}}^{b^{\prime}J^{\prime}%
}\right) \nonumber\\
&  =%
{\displaystyle\sum}
Y_{I^{\prime}}^{\prime K^{\prime}}\left(  \left(  \delta_{K^{\prime}%
}^{1^{\prime}}\delta_{1^{\prime}}^{L^{\prime}}\left(  \delta_{\overset
{.}{a^{\prime}}}^{\overset{.}{1^{\prime}}}\delta_{1^{\prime}}^{b^{\prime}%
}\overline{k}^{\nu}+\delta_{\overset{.}{a}^{\prime}}^{\overset{.}{2^{\prime}}%
}\delta_{2^{\prime}}^{b^{\prime}}\overline{k}^{e}\right)  +\delta
_{K'}^{k^{\prime}}\delta_{l^{\prime}}^{L^{\prime}}\delta_{k^{\prime}%
}^{l^{\prime}}\left(  \delta_{\overset{.}{a^{\prime}}}^{\overset{.}{1^{\prime
}}}\delta_{1^{\prime}}^{b^{\prime}}\overline{k}^{u}+\delta_{\overset{.}%
{a}^{\prime}}^{\overset{.}{2^{\prime}}}\delta_{2^{\prime}}^{b^{\prime}%
}\overline{k}^{d}\right)  \right)  Y_{L^{\prime}}^{J^{\prime}}\right.
\nonumber\\
&  \qquad\left.  -Y_{K^{\prime}}^{L^{\prime}}\left(  \left(  \delta
_{L^{\prime}}^{1^{\prime}}\delta_{1^{\prime}}^{J^{\prime}}\left(
\delta_{\overset{.}{a^{\prime}}}^{\overset{.}{1^{\prime}}}\delta_{1^{\prime}%
}^{b^{\prime}}\overline{k}^{\nu}+\delta_{\overset{.}{a}^{\prime}}^{\overset
{.}{2^{\prime}}}\delta_{2^{\prime}}^{b^{\prime}}\overline{k}^{e}\right)
+\delta_{L^{\prime}}^{l^{\prime}}\delta_{j^{\prime}}^{J^{\prime}}%
\delta_{l^{\prime}}^{j^{\prime}}\left(  \delta_{\overset{.}{a^{\prime}}%
}^{\overset{.}{1^{\prime}}}\delta_{1^{\prime}}^{b^{\prime}}\overline{k}%
^{u}+\delta_{\overset{.}{a}^{\prime}}^{\overset{.}{2^{\prime}}}\delta
_{2^{\prime}}^{b^{\prime}}\overline{k}^{d}\right)  \right)  \right)  \right)
\nonumber\\
&  =\left(  \left(  \overline{k}^{\nu}-\overline{k}^{u}\right)  \delta
_{\overset{.}{a^{\prime}}}^{\overset{.}{1^{\prime}}}\delta_{1^{\prime}%
}^{b^{\prime}}+\left(  \overline{k}^{e}-\overline{k}^{d}\right)
\delta_{\overset{.}{a^{\prime}}}^{\overset{.}{2^{\prime}}}\delta_{2^{\prime}%
}^{b^{\prime}}\right)  \Sigma_{I^{\prime}}^{J^{\prime}}%
\end{align}
where
\begin{equation}
\Sigma_{I^{\prime}}^{1^{\prime}}=%
-{\displaystyle\sum}
Y_{I^{\prime}}^{\prime k^{\prime}}Y_{k^{\prime}}^{1^{\prime}},%
\qquad
\Sigma_{I^{\prime}}^{j^{\prime}}=%
Y_{I^{\prime}}^{\prime1^{\prime}}Y_{1^{\prime}}^{j^{\prime}}
\end{equation}
Notice that if $k^{\nu}=k^{u}$ and $k^{e}=k^{d}$ which is consistent with the
picture of having the lepton number as the fourth color then $\Sigma_{I}^{J}$
will decouple. Notice that
\begin{equation}
\left(  A_{\left(  1\right)  }\right)  _{a^{\prime}I^{\prime}}^{\overset{.}%
{b}^{\prime}J^{\prime}}=\left(  \left(  k^{\nu t}-k^{ut}\right)
\delta_{a^{\prime}}^{1^{\prime}}\delta_{\overset{.}{1}^{\prime}}^{\overset
{.}{b}^{\prime}}+\left(  k^{et}-k^{dt}\right)  \delta_{a^{\prime}}^{2^{\prime
}}\delta_{\overset{.}{2}^{\prime}}^{\overset{.}{b}^{\prime}}\right)
\Sigma_{I^{\prime}}^{J^{\prime}}%
\end{equation}
which implies by the hermiticity of
\begin{equation}
A_{\overset{.}{a}^{\prime}I^{\prime}}^{b^{\prime}J^{\prime}}=\left(
A_{b^{\prime}I^{j\prime}}^{\overset{.}{a}^{\prime}I^{\prime}}\right)  ^{\ast}%
\end{equation}
that
\begin{equation}
\Sigma_{I}^{J}=\left(  \Sigma_{I}^{J}\right)  ^{\ast}%
\end{equation}
and thus belong to the $1+15$ representation of $SU\left(  4\right)  $. There
is no indication that the singlet which is equal to the trace $\Sigma_{I}^{I}$
should be absent as there is no apparent identity that equates this trace to
zero. In this case we can write%
\begin{equation}
\Sigma_{I}^{J}=\widetilde{\Sigma}_{I}^{J}+\frac{1}{4}\delta_{I}^{J}%
\,\Sigma,\qquad\Sigma=\Sigma_{I}^{I}\,,\qquad\widetilde{\Sigma}_{I}^{I}=0
\end{equation}

Thus at first order we have the Higgs fields $\phi_{a}^{\overset{.}{b}}$ and
$\Delta_{\overset{.}{a}I}$. In addition if the Yukawa couplings of the leptons
are different from the corresponding quarks (and thus requiring the breaking
of the lepton number as the fourth color) then an additional Higgs field
$\Sigma_{I}^{J}$ is also generated.

Next it is straightforward to evaluate various components of $JA_{\left(
1\right)  }J^{-1}$ which are given by
\begin{align}
\left(  JAJ^{-1}\right)  _{A}^{B} &  =\overline{A}_{A^{\prime}}^{B^{\prime}}\\
\left(  JAJ^{-1}\right)  _{A}^{B^{\prime}} &  =\overline{A}_{A^{\prime}}^{B}\\
\left(  JAJ^{-1}\right)  _{A^{\prime}}^{B^{\prime}} &  =\overline{A}_{A}^{B}\\
\left(  JAJ^{-1}\right)  _{A^{\prime}}^{B} &  =\overline{A}_{A}^{B^{\prime}}%
\end{align}
In particular
\begin{align}
\left(  JA_{\left(  1\right)  }J^{-1}\right)  _{\overset{.}{a}I}^{bJ} &
=\left(  \overline{A}_{\left(  1\right)  }\right)  _{\overset{.}{a}^{\prime
}I^{\prime}}^{b^{\prime}J^{\prime}}\nonumber\\
&  =\left(  \left(  k^{\nu}-k^{u}\right)  \delta_{\overset{.}{a}}^{\overset
{.}{1}}\delta_{1}^{b}+\left(  k^{e}-k^{d}\right)  \delta_{\overset{.}{a}%
}^{\overset{.}{2}}\delta_{2}^{b}\right)  \Sigma_{I}^{Jt}%
\end{align}%
\begin{align}
\left(  JA_{\left(  1\right)  }J^{-1}\right)  _{\overset{.}{a}I}^{\overset
{.}{b}^{\prime}J^{\prime}} &  =\left(  \overline{A}_{\left(  1\right)
}\right)  _{\overset{.}{a}^{\prime}I^{\prime}}^{\overset{.}{b}J}\nonumber\\
&  =\overline{k}^{\nu_{R}}\delta_{\overset{.}{a}}^{\overset{.}{1}}%
\delta_{1^{\prime}}^{J^{\prime}}\overline{\Delta}_{\quad I}^{\overset{.}{b}%
}\nonumber\\
&  \equiv\overline{k}^{\nu_{R}}\delta_{\overset{.}{a}}^{\overset{.}{1}}%
\delta_{1^{\prime}}^{J^{\prime}}\Delta_{\overset{.}{b}I}%
\end{align}
We now evaluate
\begin{equation}
\left(  A_{\left(  2\right)  }\right)  _{M}^{N}=%
{\displaystyle\sum}
a_{M}^{P}\left[  JA_{\left(  1\right)  }J^{-1},b\right]  _{P}^{N}%
\end{equation}
First we have
\begin{align}
\left(  A_{\left(  2\right)  }\right)  _{\alpha I}^{\beta J} &  =%
{\displaystyle\sum}
a_{\alpha I}^{\gamma K}\left(  \left(  JA_{\left(  1\right)  }J^{-1}\right)
_{\gamma K}^{\delta L}b_{\delta L}^{\beta J}-b_{\gamma K}^{\delta L}\left(
JA_{\left(  1\right)  }J^{-1}\right)  _{\delta L}^{\beta J}\right)
\nonumber\\
&  =%
{\displaystyle\sum}
X_{\alpha}^{^{\prime}\gamma}\left(  \left(  JA_{\left(  1\right)  }%
J^{-1}\right)  _{\gamma I}^{\delta J}X_{\delta}^{\beta}-X_{\gamma}^{\delta
}\left(  JA_{\left(  1\right)  }J^{-1}\right)  _{\delta I}^{\beta J}\right)
\end{align}
Thus
\begin{align}
\left(  A_{\left(  2\right)  }\right)  _{\overset{.}{a}I}^{bJ} &  =%
{\displaystyle\sum}
X_{\overset{.}{a}}^{^{\prime}\overset{.}{c}}\left(  \left(  JA_{\left(
1\right)  }J^{-1}\right)  _{\overset{.}{c}I}^{dJ}X_{d}^{b}-X_{\overset{.}{c}%
}^{\overset{.}{d}}\left(  JA_{\left(  1\right)  }J^{-1}\right)  _{\overset
{.}{d}I}^{bJ}\right)  \nonumber\\
&  =%
{\displaystyle\sum}
X_{\overset{.}{a}}^{^{\prime}\overset{.}{c}}\left(  \left(  \left(  k^{\nu
}-k^{u}\right)  \delta_{\overset{.}{c}}^{\overset{.}{1}}\delta_{1}^{d}+\left(
k^{e}-k^{d}\right)  \delta_{\overset{.}{c}}^{\overset{.}{2}}\delta_{2}%
^{d}\right)  X_{d}^{b}-\right.  \nonumber\\
&  \qquad\qquad\qquad\left.  -X_{\overset{.}{c}}^{\overset{.}{d}}\left(
\left(  k^{\nu}-k^{u}\right)  \delta_{\overset{.}{d}}^{\overset{.}{1}}%
\delta_{1}^{b}+\left(  k^{e}-k^{d}\right)  \delta_{\overset{.}{d}}%
^{\overset{.}{2}}\delta_{2}^{b}\right)  \right)  \Sigma_{I}^{Jt}\\
&  =\left(  \left(  k^{\nu}-k^{u}\right)  \left(
{\displaystyle\sum}
X_{\overset{.}{a}}^{^{\prime}\overset{.}{1}}X_{1}^{b}-X_{\overset{.}{a}%
}^{^{\prime}\overset{.}{c}}X_{\overset{.}{c}}^{\overset{.}{1}}\delta_{1}%
^{b}\right)  +\left(  k^{e}-k^{d}\right)  \left(
{\displaystyle\sum}
X_{\overset{.}{a}}^{^{\prime}\overset{.}{2}}X_{2}^{b}-X_{\overset{.}{a}%
}^{^{\prime}\overset{.}{c}}X_{\overset{.}{c}}^{\overset{.}{2}}\delta_{2}%
^{b}\right)  \right)  \Sigma_{I}^{Jt}\nonumber\\
&  =\left(  \left(  k^{\nu}-k^{u}\right)  \phi_{\overset{.}{a}}^{b}+\left(
k^{e}-k^{d}\right)  \widetilde{\phi}_{\overset{.}{a}}^{b}\right)  \Sigma
_{I}^{Jt}%
\end{align}
From the above calculation it should be clear that $\left(  A_{\left(
2\right)  }\right)  _{\overset{.}{a}I}^{bJ}$ could be expressed in terms of
the fundamental Higgs fields $\phi_{\overset{.}{a}}^{b}$ and $\Sigma_{I}^{J}$
as a consequence of the special form of the initial Dirac operator which
satisfies the order one condition for the subalgebra \eqref{subalg}. If this was not the case, then the field
$\left(  A_{\left(  2\right)  }\right)  _{\overset{.}{a}I}^{bJ}$ $\ $would be
an independent and thus fundamental Higgs field. Similarly $\left(  A_{\left(
2\right)  }\right)  _{aI}^{\overset{.}{b}J}$ is the Hermitian conjugate of
$\left(  A_{\left(  2\right)  }\right)  _{\overset{.}{b}I}^{aJ}.$. Next we
have
\begin{align}
\left(  A_{\left(  2\right)  }\right)  _{\overset{.}{a}I}^{\overset
{.}{b^{\prime}}J^{\prime}} &  =%
{\displaystyle\sum}
X_{\overset{.}{a}}^{\prime\overset{.}{c}}\left(  \left(  JA_{\left(  1\right)
}J^{-1}\right)  _{\overset{.}{c}I}^{\overset{.}{b}^{\prime}L^{\prime}%
}Y_{L^{\prime}}^{J^{\prime}}-X_{\overset{.}{c}}^{\overset{.}{d}}\left(
JA_{\left(  1\right)  }J^{-1}\right)  _{\overset{.}{d}I}^{\overset{.}%
{b}^{\prime}J^{\prime}}\right)  \nonumber\\
&  =\overline{k}^{\nu_{R}}%
{\displaystyle\sum}
\left(  X_{\overset{.}{a}}^{\prime\overset{.}{1}}Y_{1^{\prime}}^{J^{\prime}%
}-X_{\overset{.}{a}}^{\prime\overset{.}{c}}X_{\overset{.}{c}}^{\overset{.}{1}%
}\delta_{1^{\prime}}^{J^{\prime}}\right)  \overline{\Delta}_{\quad
I}^{\overset{.}{b}}\nonumber\\
&  =\overline{k}^{\nu_{R}}\Delta_{\overset{.}{a}}^{\quad J^{\prime}}%
\overline{\Delta}_{\quad I}^{\overset{.}{b}}\nonumber\\
&  =k^{\ast\nu_{R}}\Delta_{\overset{.}{a}J}\Delta_{\overset{.}{b}I}%
\end{align}

Collecting all terms we get
\begin{align}
\left(  D_{A}\right)  _{\overset{.}{a}I}^{\quad bJ}  &  =\left(
\delta_{\overset{.}{a}}^{\overset{.}{1}}\delta_{1}^{b}k^{\nu}+\delta
_{\overset{.}{a}}^{\overset{.}{2}}\delta_{2}^{b}k^{e}\right)  \delta_{I}%
^{1}\delta_{1}^{J}+\left(  \delta_{\overset{.}{a}}^{\overset{.}{1}}\delta
_{1}^{b}k^{u}+\delta_{\overset{.}{a}}^{\overset{.}{2}}\delta_{2}^{b}%
k^{d}\right)  \delta_{I}^{i}\delta_{j}^{J}\delta_{i}^{j}\nonumber\\
&  +\delta_{I}^{1}\delta_{1}^{J}\left(  k^{\nu}\phi_{\overset{.}{a}}^{b}%
+k^{e}\widetilde{\phi}_{\overset{.}{a}}^{b}\right)  +\delta_{I}^{i}\delta
_{j}^{J}\delta_{i}^{j}\left(  k^{u}\phi_{\overset{.}{a}}^{b}+k^{d}%
\widetilde{\phi}_{\overset{.}{a}}^{b}\right) \nonumber\\
&  +\left(  \left(  k^{\nu}-k^{u}\right)  \delta_{\overset{.}{a}}^{\overset
{.}{1}}\delta_{1}^{b}+\left(  k^{e}-k^{d}\right)  \delta_{\overset{.}{a}%
}^{\overset{.}{2}}\delta_{2}^{b}\right)  \Sigma_{I}^{Jt}\nonumber\\
&  +\left(  \left(  k^{\nu}-k^{u}\right)  \phi_{\overset{.}{a}}^{b}+\left(
k^{e}-k^{d}\right)  \widetilde{\phi}_{\overset{.}{a}}^{b}\right)  \Sigma
_{I}^{Jt}\nonumber\\
&  =\left(  k^{\nu}\left(  \delta_{\overset{.}{a}}^{\overset{.}{1}}\delta
_{1}^{b}+\phi_{\overset{.}{a}}^{b}\right)  +k^{e}\left(  \delta_{\overset
{.}{a}}^{\overset{.}{2}}\delta_{2}^{b}+\widetilde{\phi}_{\overset{.}{a}}%
^{b}\right)  \right)  \left(  \delta_{I}^{1}\delta_{1}^{J}+\Sigma_{I}%
^{Jt}\right) \nonumber\\
&  +\left(  k^{u}\left(  \delta_{\overset{.}{a}}^{\overset{.}{1}}\delta
_{1}^{b}+\phi_{\overset{.}{a}}^{b}\right)  +k^{d}\left(  \delta_{\overset
{.}{a}}^{\overset{.}{2}}\delta_{2}^{b}+\widetilde{\phi}_{\overset{.}{a}}%
^{b}\right)  \right)  \left(  \delta_{I}^{i}\delta_{j}^{J}\delta_{i}%
^{j}-\Sigma_{I}^{Jt}\right)
\end{align}

The other non-vanishing term is
\begin{align}
\left(  D_{A}\right)  _{\overset{.}{a}I}^{\overset{.}{\quad b}^{\prime
}J^{\prime}}  &  =k^{\ast\nu_{R}}\left(  \delta_{\overset{.}{a}}^{\overset
{.}{1}}\delta_{\overset{.}{1}^{\prime}}^{\overset{.}{b}^{\prime}}\delta
_{I}^{1}\delta_{1^{\prime}}^{J^{\prime}}+\delta_{\overset{.}{1}}^{\overset
{.}{b}}\delta_{I}^{1}\Delta_{\overset{.}{a}}^{\quad J^{\prime}}+\delta
_{\overset{.}{a}}^{\overset{.}{1}}\delta_{1^{\prime}}^{J^{\prime}}%
\overline{\Delta}_{\quad I}^{\overset{.}{b}}+\Delta_{\overset{.}{a}}^{\quad
J^{\prime}}\overline{\Delta}_{\quad I}^{\overset{.}{b}}\right) \nonumber\\
&  =k^{\ast\nu_{R}}\left(  \delta_{\overset{.}{a}}^{\overset{.}{1}}%
\delta_{1^{\prime}}^{J^{\prime}}+\Delta_{\overset{.}{a}}^{\quad J^{\prime}%
}\right)  \left(  \delta_{\overset{.}{1}}^{\overset{.}{b}}\delta_{I}%
^{1}+\overline{\Delta}_{\quad I}^{\overset{.}{b}}\right) \nonumber\\
&  \equiv k^{\ast\nu_{R}}\left(  \delta_{\overset{.}{a}}^{\overset{.}{1}%
}\delta_{J}^{1}+\Delta_{\overset{.}{a}J}\right)  \left(  \delta_{\overset
{.}{b}}^{\overset{.}{1}}\delta_{I}^{1}+\Delta_{\overset{.}{b}I}\right)
\nonumber\\
&  \equiv\left(  D_{A}\right)  _{\overset{.}{a}I\overset{.}{b}J}%
\end{align}
All other non-vanishing terms are related to the above two by Hermitian conjugation.

Note that $D_{\overset{.}{a}I}^{bJ}$ gives, after spontaneous breaking, the
Dirac masses while $D_{\overset{.}{a}I}^{\overset{.}{b}^{\prime}J^{\prime}}$
gives the Majorana masses. The Higgs fields are composite, the fundamental
ones being of similar form to those of the fermion bilinear.

It is possible to absorb the constant terms (vacuum expectation values) by
redefining the fields%
\begin{align}
\delta_{\overset{.}{a}}^{\overset{.}{1}}\delta_{1}^{b}+\phi_{\overset{.}{a}%
}^{b} &  \rightarrow\phi_{\overset{.}{a}}^{b}\\
\delta_{\overset{.}{a}}^{\overset{.}{1}}\delta_{J}^{1}+\Delta_{\overset{.}%
{a}J} &  \rightarrow\Delta_{\overset{.}{a}J}\\
\delta_{I}^{1}\delta_{1}^{J}+\Sigma_{I}^{Jt} &  \rightarrow\Sigma_{I}^{J}%
\end{align}
so that when the potential of the spectral action is minimized one will get
\begin{align}
\left\langle \phi_{\overset{.}{a}}^{b}\right\rangle  &  =\delta_{\overset
{.}{a}}^{\overset{.}{1}}\delta_{1}^{b}\\
\left\langle \Delta_{\overset{.}{a}J}\right\rangle  &  =\delta_{\overset{.}%
{a}}^{\overset{.}{1}}\delta_{J}^{1}\\
\left\langle \Sigma_{I}^{J}\right\rangle  &  =\delta_{I}^{1}\delta_{1}^{J}%
\end{align}
Thus
\begin{align}
\left(  D_{A}\right)  _{\overset{.}{a}I}^{bJ} &  =\gamma_{5}\left(  \left(
k^{\nu}\phi_{\overset{.}{a}}^{b}+k^{e}\widetilde{\phi}_{\overset{.}{a}}%
^{b}\right)  \Sigma_{I}^{J}+\left(  k^{u}\phi_{\overset{.}{a}}^{b}%
+k^{d}\widetilde{\phi}_{\overset{.}{a}}^{b}\right)  \left(  \delta_{I}%
^{J}-\Sigma_{I}^{J}\right)  \right)  \equiv\gamma_{5}\Sigma_{\overset{.}{a}%
I}^{bJ}\\
\left(  D_{A}\right)  _{\overset{.}{a}I}^{\overset{.}{b}^{\prime}J^{\prime}}
&  =\gamma_{5}k^{\ast\nu_{R}}\Delta_{\overset{.}{a}J}\Delta_{\overset{.}{b}%
I}\equiv\gamma_{5}H_{\overset{.}{a}I\overset{.}{b}J}%
\end{align}
and the fundamental Higgs fields are $\left(  2_{R},2_{L},1\right)  $,
$\left(  2_{R},1_{L},4\right)  ,$ $\left(  1_{R},1_{L},1+15\right)  .$ The
last of which $\Sigma_{I}^{J}$ drops out in the case when we take the lepton
and quark Yukawa couplings to be identical. This is a realistic possibility
and has the advantage that the Higgs sector becomes minimal. If, however, we
start with a generic initial Dirac operator, then the fields $\Sigma
_{\overset{.}{a}I}^{bJ}$ and $H_{\overset{.}{a}I\overset{.}{b}J}$ will be
independent fundamental fields in the $\left(  2_{R},2_{L},1+15\right)  $ and
$\left(  3_{R},1_{L},10\right)  $ and $\left(  1_{R},1_{L},6\right)  $
representations of $SU\left(  2\right)  _{R}\times SU\left(  2\right)
_{L}\times SU\left(  4\right)  .$

The full Dirac operator on the product space $M\times F$ \ is
\begin{equation}
\left(  D_{A}\right)  =\gamma^{\mu}D_{\mu}\otimes1+\gamma_{5}D_{F}%
\end{equation}
This gives the gauge fields
\begin{equation}
A_{\alpha I}^{\beta J}=\gamma^{\mu}%
{\displaystyle\sum}
X_{\alpha}^{\prime\gamma}\partial_{\mu}X_{\gamma}^{\beta}\delta_{I}^{J}%
\end{equation}
and in particular
\begin{align}
A_{\overset{.}{a}I}^{\overset{.}{b}J}  &  =\gamma^{\mu}%
{\displaystyle\sum}
X_{\overset{.}{a}}^{\prime\overset{.}{c}}\partial_{\mu}X_{\overset{.}{c}%
}^{\overset{.}{b}}\delta_{I}^{J}\nonumber\\
&  =\gamma^{\mu}\left(  -\frac{i}{2}g_{R}W_{\mu R}^{\alpha}\right)  \left(
\sigma^{\alpha}\right)  _{\overset{.}{a}}^{\overset{.}{b}}\delta_{I}^{J}%
\end{align}
which is the gauge field of $SU\left(  2\right)  _{R}.$ Notice that $W_{\mu
R}^{\alpha}$ are $SU\left(  2\right)  _{R}$ and not $U\left(  2\right)  $
gauge fields because $X_{\overset{.}{a}}^{\prime\overset{.}{c}}\partial_{\mu
}X_{\overset{.}{c}}^{\overset{.}{b}}$ depend on quaternionic elements.
Similarly
\begin{align}
A_{aI}^{bJ}  &  =\gamma^{\mu}%
{\displaystyle\sum}
X_{a}^{\prime c}\partial_{\mu}X_{c}^{b}\delta_{I}^{J}\nonumber\\
&  =\gamma^{\mu}\left(  -\frac{i}{2}g_{L}W_{\mu L}^{\alpha}\right)  \left(
\sigma^{\alpha}\right)  _{a}^{b}\delta_{I}^{J}%
\end{align}
where the $W_{\mu L}^{\alpha}$ are $SU\left(  2\right)  _{L}$ gauge fields. In
the conjugate sector we have
\begin{align}
A_{\alpha^{\prime}I^{\prime}}^{\beta^{\prime}J^{\prime}}  &  =\gamma^{\mu
}\delta_{\alpha^{\prime}}^{\beta^{\prime}}%
{\displaystyle\sum}
Y_{I^{\prime}}^{\prime K^{\prime}}\partial_{\mu}Y_{K^{\prime}}^{J^{\prime}%
}\nonumber\\
&  =\gamma^{\mu}\delta_{\alpha^{\prime}}^{\beta^{\prime}}\left(  \frac{i}%
{2}gV_{\mu}^{m}\left(  \lambda^{m}\right)  _{I^{\prime}}^{J^{\prime}}+\frac
{i}{2}gV_{\mu}\delta_{I^{\prime}}^{J^{\prime}}\right)
\end{align}
where $V_{\mu}^{m}$ and $V_{\mu}$ are the $U\left(  4\right)  $ gauge fields.
This implies that
\begin{align}
\left(  JAJ^{-1}\right)  _{\overset{.}{a}I}^{\overset{.}{b}J}  &
=-\gamma^{\mu}\delta_{\overset{.}{a}}^{\overset{.}{b}}\left(  \frac{i}%
{2}gV_{\mu}^{m}\left(  \lambda^{m}\right)  _{I}^{^{J}}+\frac{i}{2}gV_{\mu
}\delta_{I}^{J}\right) \\
\left(  JAJ^{-1}\right)  _{aI}^{bJ}  &  =-\gamma^{\mu}\delta_{a}^{b}\left(
\frac{i}{2}gV_{\mu}^{m}\left(  \lambda^{m}\right)  _{I}^{^{J}}+\frac{i}%
{2}gV_{\mu}\delta_{I}^{J}\right)
\end{align}
where
\begin{equation}
\mathrm{Tr}\left(  \lambda^{m}\right)  =0
\end{equation}
are the generators of the group $SU\left(  4\right)  .$ We deduce that we get
new contributions to
\begin{align}
\left(  D_{A}\right)  _{\overset{.}{a}I}^{\overset{.}{b}J}  &  =\gamma^{\mu
}\left(  D_{\mu}\delta_{\overset{.}{a}}^{\overset{.}{b}}\delta_{I}^{J}%
-\frac{i}{2}g_{R}W_{\mu R}^{\alpha}\left(  \sigma^{\alpha}\right)
_{\overset{.}{a}}^{\overset{.}{b}}\delta_{I}^{J}-\delta_{\overset{.}{a}%
}^{\overset{.}{b}}\left(  \frac{i}{2}gV_{\mu}^{m}\left(  \lambda^{m}\right)
_{I}^{^{J}}+\frac{i}{2}gV_{\mu}\delta_{I}^{J}\right)  \right) \\
\left(  D_{A}\right)  _{aI}^{bJ}  &  =\gamma^{\mu}\left(  D_{\mu}\delta
_{a}^{b}\delta_{I}^{J}-\frac{i}{2}g_{L}W_{\mu L}^{\alpha}\left(
\sigma^{\alpha}\right)  _{a}^{b}\delta_{I}^{J}-\delta_{a}^{b}\left(  \frac
{i}{2}gV_{\mu}^{m}\left(  \lambda^{m}\right)  _{I}^{^{J}}+\frac{i}{2}gV_{\mu
}\delta_{I}^{J}\right)  \right)
\end{align}
The requirement that $A$ is unimodular implies that
\begin{equation}
\mathrm{Tr}\left(  A\right)  =0
\end{equation}
which gives the condition
\begin{equation}
V_{\mu}=0
\end{equation}
and thus the gauge group of this space is
\[
SU\left(  2\right)  _{R}\times SU\left(  2\right)  _{L}\times SU\left(
4\right)
\]
Summarizing, we have%

\begin{align}
\left(  D_{A}\right)  _{\overset{.}{a}I}^{\overset{.}{b}J} &  =\gamma^{\mu
}\left(  D_{\mu}\delta_{\overset{.}{a}}^{\overset{.}{b}}\delta_{I}^{J}%
-\frac{i}{2}g_{R}W_{\mu R}^{\alpha}\left(  \sigma^{\alpha}\right)
_{\overset{.}{a}}^{\overset{.}{b}}\delta_{I}^{J}-\delta_{\overset{.}{a}%
}^{\overset{.}{b}}\left(  \frac{i}{2}gV_{\mu}^{m}\left(  \lambda^{m}\right)
_{I}^{^{J}}\right)  \right)  \otimes1_{3}\\
\left(  D_{A}\right)  _{aI}^{bJ} &  =\gamma^{\mu}\left(  D_{\mu}\delta_{a}%
^{b}\delta_{I}^{J}-\frac{i}{2}g_{L}W_{\mu L}^{\alpha}\left(  \sigma^{\alpha
}\right)  _{a}^{b}\delta_{I}^{J}-\delta_{a}^{b}\left(  \frac{i}{2}gV_{\mu}%
^{m}\left(  \lambda^{m}\right)  _{I}^{^{J}}\right)  \right)  \otimes1_{3}\\
\left(  D_{A}\right)  _{\overset{.}{a}I}^{bJ} &  =\gamma_{5}\left(  \left(
k^{\nu}\phi_{\overset{.}{a}}^{b}+k^{e}\widetilde{\phi}_{\overset{.}{a}}%
^{b}\right)  \Sigma_{I}^{J}+\left(  k^{u}\phi_{\overset{.}{a}}^{b}%
+k^{d}\widetilde{\phi}_{\overset{.}{a}}^{b}\right)  \left(  \delta_{I}%
^{J}-\Sigma_{I}^{J}\right)  \right)  \equiv\gamma_{5}\Sigma_{\overset{.}{a}%
I}^{bJ}\\
\left(  D_{A}\right)  _{\overset{.}{a}I}^{\overset{.}{b}^{\prime}J^{\prime}}
&  =\gamma_{5}k^{\ast\nu_{R}}\Delta_{\overset{.}{a}J}\Delta_{\overset{.}{b}%
I}\equiv\gamma_{5}H_{\overset{.}{a}I\overset{.}{b}J}%
\end{align}
where $1_{3}$ is for generations and
\begin{equation}
D_{\mu}=\partial_{\mu}+\frac{1}{4}\omega_{\mu}^{cd}\left(  e\right)
\gamma_{cd}%
\end{equation}
and other components are related to the ones above by
\begin{equation}
D_{A^{\prime}}^{B^{\prime}}=\overline{D}_{A}^{B},\qquad D_{A^{\prime}}%
^{B}=\overline{D}_{A}^{B^{\prime}},\qquad D_{A}^{B^{\prime}}=\overline
{D}_{A^{\prime}}^{B}.
\end{equation}
Again, for generic initial Dirac operators $\Sigma_{\overset{.}{a}I}^{bJ}$ and
$H_{\overset{.}{a}I\overset{.}{b}J}$ will be independent fundamental fields. 

We now proceed to calculate $\left(  D_{A}\right)  ^{2}.$ The first step is to
expand $D^{2}$ into the form
\begin{equation}
\left(  D_{A}\right)  ^{2}=-\left(  g^{\mu\nu}\partial_{\mu}\partial_{\nu
}+\mathcal{A}^{\mu}\partial_{\mu}+B\right)
\end{equation}
and from this extract the connection $\omega_{\mu}$
\begin{equation}
\left(  D_{A}\right)  ^{2}=-\left(  g^{\mu\nu}\nabla_{\mu}\nabla_{\nu
}+E\right)  \label{Dsquare}%
\end{equation}
where
\begin{equation}
\nabla_{\mu}=\partial_{\mu}+\mathbb{\omega}_{\mu}.
\end{equation}
This gives
\begin{align}
\mathbb{\omega}_{\mu} &  =\frac{1}{2}g_{\mu\nu}\left(  \mathcal{A}^{\nu
}+\Gamma^{\nu}\right)  \\
E &  =B-g^{\mu\nu}\left(  \partial_{\mu}\mathbb{\omega}_{\nu}+\mathbb{\omega
}_{\mu}\mathbb{\omega}_{\nu}-\Gamma_{\mu\nu}^{\rho}\mathbb{\omega}_{\rho
}\right)  \\
\Omega_{\mu\nu} &  =\partial_{\mu}\mathbb{\omega}_{\nu}-\partial_{\nu
}\mathbb{\omega}_{\mu}+\left[  \mathbb{\omega}_{\mu},\mathbb{\omega}_{\nu
}\right]
\end{align}
where $\Gamma^{\nu}=g^{\rho\sigma}\Gamma_{\rho\sigma}^{\nu}$ and $\Gamma
_{\mu\nu}^{\rho}$ is the Christoffel connection of the metric $g_{\mu\nu}.$ We
now proceed to evaluate the various components of $D^{2}:$%

\begin{align}
\left(  \left(  D_{A}\right)  ^{2}\right)  _{aI}^{bJ}  &  =\left(
D_{A}\right)  _{aI}^{\overset{.}{c}K}\left(  D_{A}\right)  _{\overset{.}{c}%
K}^{bJ}+\left(  D_{A}\right)  _{aI}^{cK}\left(  D_{A}\right)  _{cK}%
^{bJ}\nonumber\\
&  =\Sigma_{aI}^{\overset{.}{c}K}\Sigma_{\overset{.}{c}K}^{bJ}\nonumber\\
&  +\left[  \gamma^{\mu}\left(  D_{\mu}\delta_{a}^{c}\delta_{I}^{K}-\frac
{i}{2}g_{L}W_{\mu L}^{\alpha}\left(  \sigma^{\alpha}\right)  _{a}^{c}%
\delta_{I}^{K}+\delta_{a}^{c}\left(  \frac{i}{2}gV_{\mu}^{m}\left(
\lambda^{m}\right)  _{I}^{^{K}}\right)  \right)  \right. \nonumber\\
&  \left.  \gamma^{\nu}\left(  D_{\nu}\delta_{c}^{b}\delta_{K}^{J}-\frac{i}%
{2}g_{L}W_{\nu L}^{\alpha}\left(  \sigma^{\alpha}\right)  _{c}^{b}\delta
_{K}^{J}+\delta_{c}^{b}\left(  \frac{i}{2}gV_{\mu}^{m}\left(  \lambda
^{m}\right)  _{K}^{^{J}}\right)  \right)  \right]  1_{3}%
\end{align}%
\begin{align}
\left(  \left(  D_{A}\right)  ^{2}\right)  _{\overset{.}{a}I}^{\overset{.}%
{b}J}  &  =\left(  D_{A}\right)  _{\overset{.}{a}I}^{\overset{.}{c}K}\left(
D_{A}\right)  _{\overset{.}{c}K}^{\overset{.}{b}J}+\left(  D_{A}\right)
_{\overset{.}{a}I}^{cK}\left(  D_{A}\right)  _{cK}^{\overset{.}{b}J}+\left(
D_{A}\right)  _{\overset{.}{a}I}^{\overset{.}{c}^{^{\prime}}K^{^{\prime}}%
}\left(  D_{A}\right)  _{\overset{.}{c}^{^{\prime}}K^{^{\prime}}}^{\overset
{.}{b}J}\nonumber\\
&  =H_{\overset{.}{a}I\overset{.}{c}K}H^{\overset{.}{c}K\overset{.}{b}%
J}+\Sigma_{\overset{.}{a}I}^{cK}\Sigma_{cK}^{\overset{.}{b}J}\nonumber\\
&  +\left[  \gamma^{\mu}\left(  D_{\mu}\delta_{\overset{.}{a}}^{\overset{.}%
{c}}\delta_{I}^{K}-\frac{i}{2}g_{R}W_{\mu R}^{\alpha}\left(  \sigma^{\alpha
}\right)  _{\overset{.}{a}}^{\overset{.}{c}}\delta_{I}^{K}+\delta_{\overset
{.}{a}}^{\overset{.}{c}}\left(  \frac{i}{2}gV_{\mu}^{m}\left(  \lambda
^{m}\right)  _{I}^{^{K}}\right)  \right)  \right. \nonumber\\
&  \left.  \gamma^{\nu}\left(  D_{\nu}\delta_{\overset{.}{c}}^{\overset{.}{b}%
}\delta_{K}^{J}-\frac{i}{2}g_{R}W_{\nu R}^{\alpha}\left(  \sigma^{\alpha
}\right)  _{\overset{.}{c}}^{\overset{.}{b}}\delta_{K}^{J}+\delta_{\overset
{.}{c}}^{\overset{.}{b}}\left(  \frac{i}{2}gV_{\mu}^{m}\left(  \lambda
^{m}\right)  _{K}^{^{J}}\right)  \right)  \right]
\end{align}%
\begin{align}
\left(  \left(  D_{A}\right)  ^{2}\right)  _{\overset{.}{a}I}^{bJ}  &
=\left(  D_{A}\right)  _{\overset{.}{a}I}^{cK}\left(  D_{A}\right)  _{cK}%
^{bJ}+\left(  D_{A}\right)  _{\overset{.}{a}I}^{\overset{.}{c}K}\left(
D_{A}\right)  _{\overset{.}{c}K}^{bJ}\nonumber\\
&  =\gamma_{5}\gamma^{\mu}\Sigma_{\overset{.}{a}I}^{cJ}\left(  D_{\mu}%
\delta_{c}^{b}\delta_{K}^{J}-\frac{i}{2}g_{L}W_{\mu L}^{\alpha}\left(
\sigma^{\alpha}\right)  _{c}^{b}\delta_{K}^{J}+\delta_{c}^{b}\left(  \frac
{i}{2}gV_{\mu}^{m}\left(  \lambda^{m}\right)  _{K}^{^{J}}\right)  \right)
\nonumber\\
&  -\gamma_{5}\gamma^{\mu}\left(  D_{\mu}\delta_{\overset{.}{a}}^{\overset
{.}{c}}\delta_{I}^{K}-\frac{i}{2}g_{R}W_{\mu R}^{\alpha}\left(  \sigma
^{\alpha}\right)  _{\overset{.}{a}}^{\overset{.}{c}}\delta_{I}^{K}%
+\delta_{\overset{.}{a}}^{\overset{.}{c}}\left(  \frac{i}{2}gV_{\mu}%
^{m}\left(  \lambda^{m}\right)  _{I}^{^{K}}\right)  \right)  \Sigma
_{\overset{.}{c}K}^{bJ}\nonumber\\
&  =\gamma^{\mu}\gamma_{5}\nabla_{\mu}\Sigma_{\overset{.}{a}I}^{bJ}%
\end{align}
where the covariant derivative $\nabla_{\mu}$ is with respect to the gauge
group $SU\left(  2\right)  _{R}\times SU\left(  2\right)  _{L}\times SU\left(
4\right)  .$
\begin{align}
\left(  \left(  D_{A}\right)  ^{2}\right)  _{\overset{.}{a}I}^{\overset{.}%
{b}^{\prime}J^{\prime}}  &  =\left(  D_{A}\right)  _{\overset{.}{a}%
I}^{\overset{.}{c}K}\left(  D_{A}\right)  _{\overset{.}{c}K}^{\overset{.}%
{b}^{\prime}J^{\prime}}+\left(  D_{A}\right)  _{\overset{.}{a}I}^{\overset
{.}{c^{\prime}}K^{\prime}}\left(  D_{A}\right)  _{\overset{.}{c}^{\prime
}K^{\prime}}^{\overset{.}{b}^{\prime}J^{\prime}}\nonumber\\
&  =\gamma^{\mu}\gamma_{5}\left(  D_{\mu}\delta_{\overset{.}{a}}^{\overset
{.}{c}}\delta_{I}^{K}-\frac{i}{2}g_{R}W_{\mu R}^{\alpha}\left(  \sigma
^{\alpha}\right)  _{\overset{.}{a}}^{\overset{.}{c}}\delta_{I}^{K}%
+\delta_{\overset{.}{a}}^{\overset{.}{c}}\left(  \frac{i}{2}gV_{\mu}%
^{m}\left(  \lambda^{m}\right)  _{I}^{^{K}}\right)  \right)  H_{\overset{.}%
{c}K\overset{.}{b}J}\nonumber\\
&  -\gamma^{\mu}\gamma_{5}H_{\overset{.}{a}I\overset{.}{c}K}\overline{\left(
D_{\mu}\delta_{\overset{.}{c}}^{\overset{.}{b}}\delta_{K}^{J}-\frac{i}{2}%
g_{R}W_{\mu R}^{\alpha}\left(  \sigma^{\alpha}\right)  _{\overset{.}{c}%
}^{\overset{.}{b}}\delta_{K}^{J}+\delta_{\overset{.}{c}}^{\overset{.}{b}%
}\left(  \frac{i}{2}gV_{\mu}^{m}\left(  \lambda^{m}\right)  _{K}^{^{J}%
}\right)  \right)  }\nonumber\\
&  =\gamma^{\mu}\gamma_{5}\nabla_{\mu}H_{\overset{.}{a}I\overset{.}{b}J}%
\end{align}
where the covariant derivative now will be with respect to $SU\left(
2\right)  _{R}\times SU\left(  4\right)  $. Next we have%
\begin{align}
\left(  D^{2}\right)  _{\overset{.}{a}I}^{b^{\prime}J^{\prime}}  &
=D_{\overset{.}{a}I}^{\overset{.}{c}^{\prime}K^{^{\prime}}}D_{\overset{.}%
{c}^{\prime}K^{^{\prime}}}^{^{b\prime}J^{\prime}}\nonumber\\
&  =H_{\overset{.}{a}I\overset{.}{c}K}\overline{\Sigma}_{bJ}^{\overset{.}{c}K}%
\end{align}
and finally
\begin{align}
\left(  \left(  D_{A}\right)  ^{2}\right)  _{aI}^{\overset{.}{b}^{\prime
}J^{\prime}}  &  =\left(  D_{A}\right)  _{\overset{.}{a}I}^{\overset{.}{c}%
K}\left(  D_{A}\right)  _{\overset{.}{c}K}^{^{\overset{.}{b}\prime}J^{\prime}%
}\nonumber\\
&  =\Sigma_{\overset{.}{a}I}^{\overset{.}{c}K}H_{\overset{.}{c}K\overset{.}%
{b}J}%
\end{align}
We then list the entries of the matrices $\left(  \mathbb{\omega}_{\mu
}\right)  _{M}^{N}$ , $\left(  E\right)  _{M}^{N}$ which are deduced from the
form of the operator $\left(  D_{A}\right)  ^{2}.$ First we have
\begin{align}
\left(  \mathbb{\omega}_{\mu}\right)  _{aI}^{bJ}  &  =\left(  \left(  \frac
{1}{4}\omega_{\mu}^{cd}\left(  e\right)  \gamma_{cd}\right)  \delta_{a}%
^{b}\delta_{I}^{J}-\frac{i}{2}g_{L}W_{\mu L}^{\alpha}\left(  \sigma^{\alpha
}\right)  _{a}^{b}\delta_{I}^{J}-\frac{i}{2}gV_{\mu}^{m}\left(  \lambda
^{m}\right)  _{I}^{J}\delta_{a}^{b}\right)  \otimes1_{3}\\
\left(  \mathbb{\omega}_{\mu}\right)  _{\overset{.}{a}I}^{\overset{.}{b}J}  &
=\left(  \left(  \frac{1}{4}\omega_{\mu}^{cd}\left(  e\right)  \gamma
_{cd}\right)  \delta_{\overset{.}{a}}^{\overset{.}{b}}\delta_{I}^{J}-\frac
{i}{2}g_{R}W_{\mu L}^{\alpha}\left(  \sigma^{\alpha}\right)  _{\overset{.}{a}%
}^{\overset{.}{b}}\delta_{I}^{J}-\frac{i}{2}g\delta_{\overset{.}{a}}%
^{\overset{.}{b}}V_{\mu}^{m}\left(  \lambda^{m}\right)  _{I}^{J}\right)
\otimes1_{3}\\
\left(  \mathbb{\omega}_{\mu}\right)  _{A^{\prime}}^{B^{\prime}}  &  =\left(
\overline{\mathbb{\omega}}_{\mu}\right)  _{A}^{B}%
\end{align}
This in turn implies that the components of the curvature
\begin{equation}
\ \Omega_{\mu\nu}=\partial_{\mu}\omega_{\nu}-\partial_{\nu}\omega_{\mu
}+\left[  \omega_{\mu},\omega_{\nu}\right]
\end{equation}
are given by\nolinebreak\ \nolinebreak%
\begin{align}
\left(  \Omega_{\mu\nu}\right)  _{aI}^{bJ}  &  =\left(  \left(  \frac{1}%
{4}R_{\mu\nu}^{cd}\gamma_{cd}\right)  \delta_{a}^{b}\delta_{I}^{J}-\frac{i}%
{2}g_{L}W_{\mu\nu L}^{\alpha}\left(  \sigma^{\alpha}\right)  _{a}^{b}%
\delta_{I}^{J}-\frac{i}{2}gV_{\mu\nu}^{m}\left(  \lambda^{m}\right)  _{I}%
^{J}\delta_{a}^{b}\right)  \otimes1_{3}\\
\left(  \Omega_{\mu\nu}\right)  _{\overset{.}{a}I}^{\overset{.}{b}J}  &
=\left(  \left(  \frac{1}{4}R_{\mu\nu}^{cd}\gamma_{cd}\right)  \delta
_{\overset{.}{a}}^{\overset{.}{b}}\delta_{I}^{J}-\frac{i}{2}g_{R}W_{\mu\nu
R}^{\alpha}\left(  \sigma^{\alpha}\right)  _{\overset{.}{a}}^{\overset{.}{b}%
}\delta_{I}^{J}-\frac{i}{2}gV_{\mu\nu}^{m}\left(  \lambda^{m}\right)  _{I}%
^{J}\delta_{\overset{.}{a}}^{\overset{.}{b}}\right)  \otimes1_{3}\\
\left(  \Omega_{\mu\nu}\right)  _{A^{\prime}}^{B^{\prime}}  &  =\left(
\overline{\Omega}_{\mu\nu}\right)  _{A}^{B}%
\end{align}
Comparing with equation (\ref{Dsquare}) we deduce that%
\begin{align}
-\left(  E\right)  _{aI}^{bJ}  &  =\left(  \left(  \frac{1}{4}R\delta_{a}%
^{b}\delta_{I}^{J}+\frac{1}{2}\gamma^{\mu\nu}\left(  -\frac{i}{2}g_{L}%
W_{\mu\nu L}^{\alpha}\left(  \sigma^{\alpha}\right)  _{a}^{b}\delta_{I}%
^{J}-\frac{i}{2}gV_{\mu\nu}^{m}\left(  \lambda^{m}\right)  _{I}^{J}\delta
_{a}^{b}\right)  \right)  1_{3}+\Sigma_{aI}^{\overset{.}{c}K}\Sigma
_{\overset{.}{c}K}^{bJ}\right) \\
\left(  -E\right)  _{\overset{.}{a}I}^{\overset{.}{b}J}  &  =\left(  \left(
\frac{1}{4}R\delta_{\overset{.}{a}}^{\overset{.}{b}}\delta_{I}^{J}+\frac{1}%
{2}\gamma^{\mu\nu}\left(  -\frac{i}{2}g_{R}W_{\mu\nu R}^{\alpha}\left(
\sigma^{\alpha}\right)  _{\overset{.}{a}}^{\overset{.}{b}}\delta_{I}^{J}%
-\frac{i}{2}gV_{\mu\nu}^{m}\left(  \lambda^{m}\right)  _{I}^{J}\delta
_{\overset{.}{a}}^{\overset{.}{b}}\right)  \right)  1_{3}\right. \\
&  \qquad\qquad\left.  +H_{\overset{.}{a}I\overset{.}{c}K}H^{\overset{.}%
{c}K\overset{.}{b}J}+\Sigma_{\overset{.}{a}I}^{cK}\Sigma_{cK}^{\overset{.}%
{b}J}\right) \\
-\left(  E\right)  _{\overset{.}{a}I}^{bJ}  &  =\gamma^{\mu}\gamma_{5}%
\nabla_{\mu}\Sigma_{\overset{.}{a}I}^{bJ}\\
-\left(  E\right)  _{\overset{.}{a}I}^{\overset{.}{b}^{\prime}J^{\prime}}  &
=\gamma^{\mu}\gamma_{5}\nabla_{\mu}H_{\overset{.}{a}I\overset{.}{b}J}\\
\left(  -E\right)  _{\overset{.}{a}I}^{b^{\prime}J^{\prime}}  &
=H_{\overset{.}{a}I\overset{.}{c}K}\overline{\Sigma}_{bJ}^{\overset{.}{c}K}\\
\left(  -E\right)  _{aI}^{\overset{.}{b}^{\prime}J^{\prime}}  &  =\Sigma
_{aI}^{\overset{.}{c}K}H_{\overset{.}{c}K\overset{.}{b}J}%
\end{align}
Evaluating the various traces of the $384\times384$ matrices on spinor and
generation space, we get
\begin{equation}
\text{\textrm{Tr} }\left(  E\right)  =\text{\textrm{tr} }\left(  E_{A}%
^{A}+E_{A^{\prime}}^{A^{\prime}}\right)  =\text{\textrm{tr} }\left(  E_{A}%
^{A}+\overline{E}_{A}^{A}\right)
\end{equation}%
\begin{align}
-\text{\textrm{tr} }\left(  E\right)  _{aI}^{aI}  &  =4\left[  \frac{3}%
{4}R\left(  2\right)  \left(  4\right)  +H_{aI}^{\overset{.}{cK}}%
H_{\overset{.}{c}K}^{aI}\right] \\
-\text{\textrm{tr} }\left(  E\right)  _{\overset{.}{a}I}^{\overset{.}{a}J}  &
=4\left[  \frac{3}{4}R\left(  2\right)  \left(  4\right)  +H_{\overset{.}%
{a}I\overset{.}{c}K}H^{\overset{.}{c}K\overset{.}{a}I}+\Sigma_{\overset{.}%
{a}I}^{cK}\Sigma_{cK}^{\overset{.}{a}I}\right]
\end{align}%
\begin{equation}
-\frac{1}{2}\text{\textrm{Tr} }\left(  E\right)  =4\left(  12R+H_{\overset
{.}{a}I\overset{.}{c}K}H^{\overset{.}{c}K\overset{.}{a}I}+2\Sigma_{\overset
{.}{a}I}^{cK}\Sigma_{cK}^{\overset{.}{a}I}\right)
\end{equation}
Next
\begin{align}
\text{\textrm{Tr }}\left(  \Omega_{\mu\nu}^{2}\right)  _{M}^{M}  &
=2\text{\textrm{Tr} }\left(  \Omega_{\mu\nu}^{2}\right)  _{A}^{A}\nonumber\\
&  =2\text{\textrm{Tr}}\left(  \left(  \Omega_{\mu\nu}^{2}\right)
_{\overset{.}{a}I}^{\overset{.}{a}I}+\left(  \Omega_{\mu\nu}^{2}\right)
_{aI}^{aI}\right)
\end{align}%
\begin{align}
\text{\textrm{Tr}}\left(  \Omega_{\mu\nu}^{2}\right)  _{aI}^{aI}  &
=\text{\textrm{Tr}}\left(  \left(  \left(  \frac{1}{4}R_{\mu\nu}^{cd}%
\gamma_{cd}\right)  \delta_{a}^{b}\delta_{I}^{J}-\frac{i}{2}g_{L}W_{\mu\nu
L}^{\alpha}\left(  \sigma^{\alpha}\right)  _{a}^{b}\delta_{I}^{J}-\frac{i}%
{2}gV_{\mu\nu}^{m}\left(  \lambda^{m}\right)  _{i}^{j}\delta_{a}^{b}\right)
\otimes1_{3}\right)  ^{2}\nonumber\\
&  =4\left[  -\frac{1}{8}R_{\mu\nu\rho\sigma}^{2}\left(  4\right)  \left(
2\right)  \left(  3\right)  -\frac{1}{4}g_{L}^{2}\left(  W_{\mu\nu}^{\alpha
}\right)  ^{2}\left(  4\right)  \left(  2\right)  \left(  3\right)  -\frac
{1}{4}g^{2}\left(  V_{\mu\nu}^{m}\right)  ^{2}\left(  3\right)  \left(
2\right)  \left(  2\right)  \right] \nonumber\\
&  =4\left[  -3R_{\mu\nu\rho\sigma}^{2}-6g_{L}^{2}\left(  W_{\mu\nu L}%
^{\alpha}\right)  ^{2}-3g^{2}\left(  V_{\mu\nu}^{m}\right)  ^{2}\right] \\
\text{\textrm{Tr}}\left(  \Omega_{\mu\nu}^{2}\right)  _{\overset{.}{a}%
I}^{\overset{.}{a}I}  &  =\text{\textrm{Tr}}\left(  \left(  \left(  \frac
{1}{4}R_{\mu\nu}^{cd}\gamma_{cd}\right)  \delta_{\overset{.}{a}}^{\overset
{.}{b}}\delta_{I}^{J}-\frac{i}{2}g_{R}W_{\mu\nu R}^{\alpha}\left(
\sigma^{\alpha}\right)  _{\overset{.}{a}}^{\overset{.}{b}}\delta_{I}^{J}%
-\frac{i}{2}gV_{\mu\nu}^{m}\left(  \lambda^{m}\right)  _{I}^{J}\delta
_{\overset{.}{a}}^{\overset{.}{b}}\right)  \otimes1_{3}\right)  ^{2}%
\nonumber\\
&  =4\left[  -3R_{\mu\nu\rho\sigma}^{2}-6g_{R}^{2}\left(  W_{\mu\nu R}%
^{\alpha}\right)  ^{2}-3g^{2}\left(  V_{\mu\nu}^{m}\right)  ^{2}\right]
\end{align}
Therefore
\begin{equation}
\frac{1}{2}\text{\textrm{Tr} }\left(  \Omega_{\mu\nu}^{2}\right)  _{M}%
^{M}=24\left[  -R_{\mu\nu\rho\sigma}^{2}-g_{L}^{2}\left(  W_{\mu\nu L}%
^{\alpha}\right)  ^{2}-g_{R}^{2}\left(  W_{\mu\nu R}^{\alpha}\right)
^{2}-g^{2}\left(  V_{\mu\nu}^{m}\right)  ^{2}\right]
\end{equation}
Next we compute \nolinebreak%
\begin{equation}
\left(  E^{2}\right)  _{A}^{B}=E_{A}^{C}E_{C}^{B}+E_{A}^{C^{\prime}%
}E_{C^{\prime}}^{B}%
\end{equation}
and listing the components of this matrix we get
\begin{equation}
\left(  E^{2}\right)  _{aI}^{bJ}=E_{aI}^{cK}E_{cK}^{bj}+E_{aI}^{\overset{.}%
{c}K}E_{\overset{.}{c}K}^{bJ}+E_{aI}^{\overset{.}{c}^{\prime}K^{\prime}%
}E_{\overset{.}{c}^{\prime}K^{\prime}}^{bJ}%
\end{equation}%
\begin{equation}
\left(  E^{2}\right)  _{\overset{.}{a}I}^{\overset{.}{b}J}=E_{\overset{.}{a}%
I}^{\overset{.}{c}K}E_{\overset{.}{c}K}^{\overset{.}{b}J}+E_{\overset{.}{a}%
I}^{cK}E_{cK}^{\overset{.}{b}J}+E_{\overset{.}{a}I}^{c^{\prime}K^{\prime}%
}E_{c^{\prime}K^{\prime}}^{\overset{.}{b}J}+E_{\overset{.}{a}I}^{\overset
{.}{c}^{^{\prime}}K^{^{\prime}}}E_{\overset{.}{c}^{^{\prime}}K^{^{\prime}}%
}^{\overset{.}{b}J}%
\end{equation}
Collecting terms and tracing we obtain for the right-handed components%
\begin{align}
\text{\textrm{tr }}\left(  E^{2}\right)  _{\overset{.}{a}I}^{\overset{.}{a}I}
&  =\text{\textrm{tr}}\left\{  \left(  \gamma^{\mu}\gamma_{5}\nabla_{\mu
}\Sigma_{\overset{.}{a}I}^{bJ}\gamma^{\nu}\gamma_{5}\nabla_{\nu}\Sigma
_{bJ}^{\overset{.}{a}I}\right)  +\left(  \gamma^{\mu}\gamma_{5}\nabla_{\mu
}H_{\overset{.}{a}I\overset{.}{b}J}\gamma^{\nu}\gamma_{5}\nabla_{\nu
}H^{\overset{.}{a}I\overset{.}{b}J}\right)  +H_{\overset{.}{a}I\overset{.}%
{c}K}\Sigma_{bJ}^{\overset{.}{c}K}H^{\overset{.}{a}I\overset{.}{d}L}%
\Sigma_{\overset{.}{d}L}^{bJ}\right. \nonumber\\
&  +\left(  \left(  \frac{1}{4}R\delta_{\overset{.}{a}}^{\overset{.}{b}}%
\delta_{I}^{J}+\frac{1}{2}\gamma^{\mu\nu}\left(  -\frac{i}{2}g_{R}W_{\mu\nu
R}^{\alpha}\left(  \sigma^{\alpha}\right)  _{\overset{.}{a}}^{\overset{.}{b}%
}\delta_{I}^{J}-\frac{i}{2}gV_{\mu\nu}^{m}\left(  \lambda^{m}\right)  _{I}%
^{J}\delta_{\overset{.}{a}}^{\overset{.}{b}}\right)  \right)  1_{3}\right.
\nonumber\\
&  \qquad\left.  \left.  +H_{\overset{.}{a}I\overset{.}{c}K}H^{\overset{.}%
{c}K\overset{.}{b}J}+\Sigma_{\overset{.}{a}I}^{cK}\Sigma_{cK}^{\overset{.}%
{b}J}\right)  ^{2}\right\} \\
&  =4\left[  \frac{1}{4}\left(  -2\right)  \left(  -\frac{1}{4}g_{R}%
^{2}\left(  W_{\mu\nu R}^{\alpha}\right)  ^{2}\left(  2\right)  \left(
4\right)  (3)-\frac{1}{4}g^{2}\left(  V_{\mu\nu}^{m}\right)  ^{2}\left(
2\right)  \left(  2\right)  \left(  3\right)  \right)  +\frac{1}{16}%
R^{2}\left(  2\right)  \left(  4\right)  \left(  3\right)  \right. \nonumber\\
&  +\frac{1}{2}R\left(  H_{\overset{.}{a}I\overset{.}{c}K}H^{\overset{.}%
{c}K\overset{.}{a}I}+\Sigma_{\overset{.}{a}I}^{cK}\Sigma_{cK}^{\overset{.}%
{a}I}\right)  +\nabla_{\mu}H_{\overset{.}{a}I\overset{.}{b}J}\nabla^{\mu
}H^{\overset{.}{a}I\overset{.}{b}J}+\nabla_{\mu}\Sigma_{\overset{.}{a}I}%
^{bJ}\nabla^{\mu}\Sigma_{bJ}^{\overset{.}{a}I}\nonumber\\
&  \left.  +H_{\overset{.}{a}I\overset{.}{c}K}\Sigma_{bJ}^{\overset{.}{c}%
K}H^{\overset{.}{a}I\overset{.}{d}L}\Sigma_{\overset{.}{d}L}^{bJ}+\left\vert
H_{\overset{.}{a}I\overset{.}{c}K}H^{\overset{.}{c}K\overset{.}{b}J}%
+\Sigma_{\overset{.}{a}I}^{cK}\Sigma_{cK}^{\overset{.}{b}J}\right\vert
^{2}\right] \nonumber\\
&  =4\left[  \frac{3}{2}\left(  2g_{R}^{2}\left(  W_{\mu\nu R}^{\alpha
}\right)  ^{2}+g_{3}^{2}\left(  V_{\mu\nu}^{m}\right)  ^{2}\right)  +\frac
{3}{2}R^{2}+\nabla_{\mu}H_{\overset{.}{a}I\overset{.}{b}J}\nabla^{\mu
}H^{\overset{.}{a}I\overset{.}{b}J}+\nabla_{\mu}\Sigma_{\overset{.}{a}I}%
^{bJ}\nabla^{\mu}\Sigma_{bJ}^{\overset{.}{a}I}+\right. \nonumber\\
&  \left.  +\frac{1}{2}R\left(  H_{\overset{.}{a}I\overset{.}{c}K}%
H^{\overset{.}{c}K\overset{.}{a}I}+\Sigma_{\overset{.}{a}I}^{cK}\Sigma
_{cK}^{\overset{.}{a}I}\right)  +H_{\overset{.}{a}I\overset{.}{c}K}\Sigma
_{bJ}^{\overset{.}{c}K}H^{\overset{.}{a}I\overset{.}{d}L}\Sigma_{\overset
{.}{d}L}^{bJ}+\left\vert H_{\overset{.}{a}I\overset{.}{c}K}H^{\overset{.}%
{c}K\overset{.}{b}J}+\Sigma_{\overset{.}{a}I}^{cK}\Sigma_{cK}^{\overset{.}%
{b}J}\right\vert ^{2}\right]
\end{align}
and for the left-handed components%
\begin{align}
\text{\textrm{tr} }\left(  E^{2}\right)  _{aI}^{aI}  &  =\text{\textrm{tr}%
}\left\{  \left(  \left(  \frac{R}{4}\delta_{a}^{b}\delta_{I}^{J}+\frac{1}%
{2}\gamma^{\mu\nu}\left(  -\frac{i}{2}g_{L}W_{\mu\nu L}^{\alpha}\left(
\sigma^{\alpha}\right)  _{a}^{b}\delta_{I}^{J}-\frac{i}{2}gV_{\mu\nu}%
^{m}\left(  \lambda^{m}\right)  _{I}^{J}\right)  \delta_{a}^{b}\right)
1_{3}+\Sigma_{aI}^{\overset{.}{cK}}\Sigma_{\overset{.}{c}K}^{bJ}\right)
^{2}\right. \nonumber\\
&  \left.  +\gamma^{\mu}\gamma_{5}\nabla_{\mu}\Sigma_{aI}^{\overset{.}{c}%
K}\gamma^{\nu}\gamma_{5}\nabla_{\nu}\Sigma_{\overset{.}{c}K}^{aI}+\left\vert
\Sigma_{aI}^{\overset{.}{c}K}H_{\overset{.}{c}I\overset{.}{b}J}\right\vert
^{2}\right\} \nonumber\\
&  =4\left[  \frac{1}{4}\left(  -2\right)  \left(  -\frac{1}{4}g_{L}%
^{2}\left(  W_{\mu\nu L}^{\alpha}\right)  ^{2}\left(  2\right)  \left(
4\right)  (3)-\frac{1}{4}g_{3}^{2}\left(  V_{\mu\nu}^{m}\right)  ^{2}\left(
2\right)  \left(  2\right)  \left(  3\right)  \right)  +\frac{1}{16}%
R^{2}\left(  2\right)  \left(  4\right)  \left(  3\right)  \right. \nonumber\\
&  \left.  +\frac{1}{2}R\Sigma_{aI}^{\overset{.}{c}K}\Sigma_{\overset{.}{c}%
K}^{aI}+\nabla_{\mu}\Sigma_{aI}^{\overset{.}{c}K}\nabla^{\mu}\Sigma
_{\overset{.}{c}K}^{aI}+\Sigma_{aI}^{\overset{.}{c}K}\Sigma_{\overset{.}{c}%
K}^{bJ}\Sigma_{bJ}^{\overset{.}{dL}}\Sigma_{\overset{.}{d}L}^{aI}+\left\vert
\Sigma_{aI}^{\overset{.}{c}K}H_{\overset{.}{c}I\overset{.}{b}J}\right\vert
^{2}\right] \nonumber\\
&  =4\left[  \frac{3}{2}\left(  2g_{L}^{2}\left(  W_{\mu\nu L}^{\alpha
}\right)  ^{2}+g_{3}^{2}\left(  V_{\mu\nu}^{m}\right)  ^{2}\right)  +\frac
{3}{2}R^{2}+\nabla_{\mu}\Sigma_{aI}^{\overset{.}{c}K}\nabla^{\mu}%
\Sigma_{\overset{.}{c}K}^{aI}\right. \nonumber\\
&  \qquad\left.  +\frac{1}{2}R\Sigma_{aI}^{\overset{.}{c}K}\Sigma_{\overset
{.}{c}K}^{aI}+\Sigma_{aI}^{\overset{.}{c}K}\Sigma_{\overset{.}{c}K}^{bJ}%
\Sigma_{bJ}^{\overset{.}{dL}}\Sigma_{\overset{.}{d}L}^{aI}+\left\vert
\Sigma_{aI}^{\overset{.}{c}K}H_{\overset{.}{c}I\overset{.}{b}J}\right\vert
^{2}\right]
\end{align}
Collecting all terms we finally get%
\begin{align}
\frac{1}{2}\text{\textrm{tr} }\left(  E^{2}\right)   &  =4\left[  3\left(
g_{L}^{2}\left(  W_{\mu\nu L}^{\alpha}\right)  ^{2}+g^{2}\left(  V_{\mu\nu
}^{m}\right)  ^{2}+g_{R}^{2}\left(  W_{\mu\nu R}^{\alpha}\right)  ^{2}%
+R^{2}\right)  \right. \nonumber\\
&  +2\nabla_{\mu}\Sigma_{aI}^{\overset{.}{c}K}\nabla^{\mu}\Sigma_{\overset
{.}{c}K}^{aI}+\nabla_{\mu}H_{\overset{.}{a}I\overset{.}{b}J}\nabla^{\mu
}H^{\overset{.}{a}I\overset{.}{b}J}+\frac{1}{2}R\left(  H_{\overset{.}%
{a}I\overset{.}{c}K}H^{\overset{.}{c}K\overset{.}{a}I}+2\Sigma_{\overset{.}%
{a}I}^{cK}\Sigma_{cK}^{\overset{.}{a}I}\right) \nonumber\\
&  \left.  +2\Sigma_{aI}^{\overset{.}{c}K}\Sigma_{\overset{.}{c}K}^{bJ}%
\Sigma_{bJ}^{\overset{.}{dL}}\Sigma_{\overset{.}{d}L}^{aI}+4H_{\overset{.}%
{a}I\overset{.}{c}K}\Sigma_{bJ}^{\overset{.}{c}K}H^{\overset{.}{a}I\overset
{.}{d}L}\Sigma_{\overset{.}{d}L}^{bJ}+\left\vert H_{\overset{.}{a}I\overset
{.}{c}K}H^{\overset{.}{c}K\overset{.}{b}J}\right\vert ^{2}\right]
\end{align}
The first two Seely-de Witt coefficients are, first for $a_{0}$
\begin{align}
a_{0}  &  =\frac{1}{16\pi^{2}}%
{\displaystyle\int}
d^{4}x\sqrt{g}\text{Tr}\left(  1\right) \nonumber\\
&  =\frac{1}{16\pi^{2}}\left(  4\right)  \left(  32\right)  \left(  3\right)
{\displaystyle\int}
d^{4}x\sqrt{g}\nonumber\\
&  =\frac{24}{\pi^{2}}%
{\displaystyle\int}
d^{4}x\sqrt{g}%
\end{align}
then for $a_{2}:$%
\begin{align}
a_{2}  &  =\frac{1}{16\pi^{2}}%
{\displaystyle\int}
d^{4}x\sqrt{g}\text{\textrm{Tr}}\left(  E+\frac{1}{6}R\right) \nonumber\\
&  =\frac{1}{16\pi^{2}}%
{\displaystyle\int}
d^{4}x\sqrt{g}\left(  \left(  R(-96+64\right)  -8\left(  H_{\overset{.}%
{a}I\overset{.}{c}K}H^{\overset{.}{c}K\overset{.}{a}I}+2\Sigma_{\overset{.}%
{a}I}^{cK}\Sigma_{cK}^{\overset{.}{a}I}\right)  \right) \nonumber\\
&  =-\frac{2}{\pi^{2}}%
{\displaystyle\int}
d^{4}x\sqrt{g}\left(  R+\frac{1}{4}\left(  H_{\overset{.}{a}I\overset{.}{c}%
K}H^{\overset{.}{c}K\overset{.}{a}I}+2\Sigma_{\overset{.}{a}I}^{cK}\Sigma
_{cK}^{\overset{.}{a}I}\right)  \right)
\end{align}
With all the above information we can now compute the Seeley-de Witt
coefficient $a_{4}:$%
\begin{equation}
a_{4}=\frac{1}{16\pi^{2}}%
{\displaystyle\int}
d^{4}x\sqrt{g}\text{\textrm{Tr}}\left(  \frac{1}{360}\left(  5R^{2}-2R_{\mu
\nu}^{2}+2R_{\mu\nu\rho\sigma}^{2}\right)  1+\frac{1}{2}\left(  E^{2}+\frac
{1}{3}RE+\frac{1}{6}\Omega_{\mu\nu}^{2}\right)  \right)
\end{equation}
and where we\ have omitted the surface terms. Thus%
\begin{align}
&  \frac{1}{2}\text{\textrm{Tr}}\left(  E^{2}+\frac{1}{3}RE+\frac{1}{6}%
\Omega_{\mu\nu}^{2}\right) \nonumber\\
&  =4\left[  3\left(  g_{L}^{2}\left(  W_{\mu\nu L}^{\alpha}\right)
^{2}+g^{2}\left(  V_{\mu\nu}^{m}\right)  ^{2}+g_{R}^{2}\left(  W_{\mu\nu
R}^{\alpha}\right)  ^{2}+R^{2}\right)  +2\nabla_{\mu}\Sigma_{aI}^{\overset
{.}{c}K}\nabla^{\mu}\Sigma_{\overset{.}{c}K}^{aI}\right. \nonumber\\
&  +\nabla_{\mu}H_{\overset{.}{a}I\overset{.}{b}J}\nabla^{\mu}H^{\overset
{.}{a}I\overset{.}{b}J}+\frac{1}{2}R\left(  H_{\overset{.}{a}I\overset{.}{c}%
K}H^{\overset{.}{c}K\overset{.}{a}I}+2\Sigma_{\overset{.}{a}I}^{cK}\Sigma
_{cK}^{\overset{.}{a}I}\right) \nonumber\\
&  +4H_{\overset{.}{a}I\overset{.}{c}K}\Sigma_{bJ}^{\overset{.}{c}%
K}H^{\overset{.}{a}I\overset{.}{d}L}\Sigma_{\overset{.}{d}L}^{bJ}+2\Sigma
_{aI}^{\overset{.}{c}K}\Sigma_{\overset{.}{c}K}^{bJ}\Sigma_{bJ}^{\overset
{.}{dL}}\Sigma_{\overset{.}{d}L}^{aI}+\left\vert H_{\overset{.}{a}I\overset
{.}{c}K}H^{\overset{.}{c}K\overset{.}{b}J}\right\vert ^{2}\nonumber\\
&  -\frac{1}{3}R\left(  12R+H_{\overset{.}{a}I\overset{.}{c}K}H^{\overset
{.}{c}K\overset{.}{a}I}+2\Sigma_{\overset{.}{a}I}^{cK}\Sigma_{cK}^{\overset
{.}{a}I}\right) \nonumber\\
&  \left.  -R_{\mu\nu\rho\sigma}^{2}-g_{L}^{2}\left(  W_{\mu\nu L}^{\alpha
}\right)  ^{2}-g_{R}^{2}\left(  W_{\mu\nu R}^{\alpha}\right)  ^{2}%
-g^{2}\left(  V_{\mu\nu}^{m}\right)  ^{2}\right] \nonumber\\
&  =4\left[  -R_{\mu\nu\rho\sigma}^{2}-R^{2}+2g_{L}^{2}\left(  W_{\mu\nu
L}^{\alpha}\right)  ^{2}+2g_{R}^{2}\left(  W_{\mu\nu R}^{\alpha}\right)
^{2}+2g^{2}\left(  V_{\mu\nu}^{m}\right)  ^{2}\right. \nonumber\\
&  +2\nabla_{\mu}\Sigma_{aI}^{\overset{.}{c}K}\nabla^{\mu}\Sigma_{\overset
{.}{c}K}^{aI}+\nabla_{\mu}H_{\overset{.}{a}I\overset{.}{b}J}\nabla^{\mu
}H^{\overset{.}{a}I\overset{.}{b}J}+\frac{1}{6}R\left(  H_{\overset{.}%
{a}I\overset{.}{c}K}H^{\overset{.}{c}K\overset{.}{a}I}+2\Sigma_{\overset{.}%
{a}I}^{cK}\Sigma_{cK}^{\overset{.}{a}I}\right) \nonumber\\
&  \left.  +\left\vert H_{\overset{.}{a}I\overset{.}{c}K}H^{\overset{.}%
{c}K\overset{.}{b}J}\right\vert ^{2}+4H_{\overset{.}{a}I\overset{.}{c}K}%
\Sigma_{bJ}^{\overset{.}{c}K}H^{\overset{.}{a}I\overset{.}{d}L}\Sigma
_{\overset{.}{d}L}^{bJ}+2\Sigma_{aI}^{\overset{.}{c}K}\Sigma_{\overset{.}{c}%
K}^{bJ}\Sigma_{bJ}^{\overset{.}{dL}}\Sigma_{\overset{.}{d}L}^{aI}\right]
\end{align}
Collecting terms we get
\begin{align}
a_{4}  &  =\frac{1}{2\pi^{2}}%
{\displaystyle\int}
d^{4}x\sqrt{g}\left[  \frac{1}{30}\left(  5R^{2}-8R_{\mu\nu}^{2}-7R_{\mu
\nu\rho\sigma}^{2}\right)  +g_{L}^{2}\left(  W_{\mu\nu L}^{\alpha}\right)
^{2}+g_{R}^{2}\left(  W_{\mu\nu R}^{\alpha}\right)  ^{2}+g^{2}\left(
V_{\mu\nu}^{m}\right)  ^{2}\right. \nonumber\\
&  \qquad\qquad+\ \nabla_{\mu}\Sigma_{aI}^{\overset{.}{c}K}\nabla^{\mu}%
\Sigma_{\overset{.}{c}K}^{aI}+\frac{1}{2}\nabla_{\mu}H_{\overset{.}%
{a}I\overset{.}{b}J}\nabla^{\mu}H^{\overset{.}{a}I\overset{.}{b}J}+\frac
{1}{12}R\left(  H_{\overset{.}{a}I\overset{.}{c}K}H^{\overset{.}{c}%
K\overset{.}{a}I}+\Sigma_{\overset{.}{a}I}^{cK}\Sigma_{cK}^{\overset{.}{a}%
I}+H_{aI}^{\overset{.}{c}K}H_{\overset{.}{c}K}^{aI}\right) \nonumber\\
&  \qquad\qquad\left.  +\frac{1}{2}\left\vert H_{\overset{.}{a}I\overset{.}%
{c}K}H^{\overset{.}{c}K\overset{.}{b}J}\right\vert ^{2}+2H_{\overset{.}%
{a}I\overset{.}{c}K}\Sigma_{bJ}^{\overset{.}{c}K}H^{\overset{.}{a}I\overset
{.}{d}L}\Sigma_{dL}^{bJ}+\Sigma_{aI}^{\overset{.}{c}K}\Sigma_{\overset{.}{c}%
K}^{bJ}\Sigma_{bJ}^{\overset{.}{dL}}\Sigma_{\overset{.}{d}L}^{aI}\right]
\end{align}
Using the identities%
\begin{align}
R_{\mu\nu\rho\sigma}^{2}  &  =2C_{\mu\nu\rho\sigma}^{2}+\frac{1}{3}%
R^{2}-R^{\ast}R^{\ast}\\
R_{\mu\nu}^{2}  &  =\frac{1}{2}C_{\mu\nu\rho\sigma}^{2}+\frac{1}{3}R^{2}%
-\frac{1}{2}R^{\ast}R^{\ast}%
\end{align}
where\ $R^{\ast}R^{\ast}=\frac{1}{4}\epsilon^{\mu\nu\rho\sigma}\epsilon
_{\alpha\beta\gamma\delta}R_{\mu\nu}^{\quad\alpha\beta}R_{\rho\sigma}%
^{\quad\gamma\delta}.$
\begin{align}
\frac{1}{30}\left(  5R^{2}-8R_{\mu\nu}^{2}-7R_{\mu\nu\rho\sigma}^{2}\right)
&  =R^{2}\frac{1}{30}\left(  5-\frac{8}{3}-\frac{7}{3}\right)  +\frac{1}%
{30}C_{\mu\nu\rho\sigma}^{2}\left(  -4-14\right)  +\frac{1}{30}R^{\ast}%
R^{\ast}\left(  4+7\right) \nonumber\\
&  =-\frac{3}{5}C_{\mu\nu\rho\sigma}^{2}+\frac{11}{30}R^{\ast}R^{\ast}%
\end{align}
Then $a_{4}$ simplifies to
\begin{align}
a_{4}  &  =\frac{1}{2\pi^{2}}%
{\displaystyle\int}
d^{4}x\sqrt{g}\left[  -\frac{3}{5}C_{\mu\nu\rho\sigma}^{2}+\frac{11}%
{30}R^{\ast}R^{\ast}+g_{L}^{2}\left(  W_{\mu\nu L}^{\alpha}\right)  ^{2}%
+g_{R}^{2}\left(  W_{\mu\nu R}^{\alpha}\right)  ^{2}+g^{2}\left(  V_{\mu\nu
}^{m}\right)  ^{2}\right. \nonumber\\
&  \qquad\qquad+\nabla_{\mu}\Sigma_{aI}^{\overset{.}{c}K}\nabla^{\mu}%
\Sigma_{\overset{.}{c}K}^{aI}+\frac{1}{2}\nabla_{\mu}H_{\overset{.}%
{a}I\overset{.}{b}J}\nabla^{\mu}H^{\overset{.}{a}I\overset{.}{b}J}+\frac
{1}{12}R\left(  H_{\overset{.}{a}I\overset{.}{c}K}H^{\overset{.}{c}%
K\overset{.}{a}I}+2\Sigma_{\overset{.}{a}I}^{cK}\Sigma_{cK}^{\overset{.}{a}%
I}\right) \nonumber\\
&  \qquad\qquad\left.  +\frac{1}{2}\left\vert H_{\overset{.}{a}I\overset{.}%
{c}K}H^{\overset{.}{c}K\overset{.}{b}J}\right\vert ^{2}+2H_{\overset{.}%
{a}I\overset{.}{c}K}\Sigma_{bJ}^{\overset{.}{c}K}H^{\overset{.}{a}I\overset
{.}{d}L}\Sigma_{\overset{.}{d}L}^{bJ}+\Sigma_{aI}^{\overset{.}{c}K}%
\Sigma_{\overset{.}{c}K}^{bJ}\Sigma_{bJ}^{\overset{.}{dL}}\Sigma_{\overset
{.}{d}L}^{aI}\right]
\end{align}

\bigskip

\begin{acknowledgments}
AHC is supported in part by the National Science Foundation under Grant No.
Phys-0854779 and Phys-1202671. WDvS thanks IH\'{E}S for hospitality during a
visit from January-March 2013.
\end{acknowledgments}

\end{document}